\documentclass[11pt]{article}

\pdfoutput=1
\usepackage{amsmath,amssymb}
\usepackage{graphicx}
\usepackage[english]{babel}
\usepackage[utf8x]{inputenc}
\usepackage[T1]{fontenc}
\usepackage{slashed}
\usepackage{cite}

\usepackage[colorinlistoftodos]{todonotes}
\usepackage[colorlinks=true,allcolors=blue]{hyperref}
\newcommand{\eref}[1]{Eq.~(\ref{#1})}
\newcommand{\figref}[1]{Fig.~\ref{#1}}
\newcommand{\tabref}[1]{Tab.~\ref{#1}}
\newcommand{\secref}[1]{Sec.~\ref{#1}}

\def\ltap{\ \raise.3ex\hbox{$<$\kern-.75em\lower1ex\hbox{$\sim$}}\ }
\def\gtap{\ \raise.3ex\hbox{$>$\kern-.75em\lower1ex\hbox{$\sim$}}\ }

\newcommand{\ra}{\rightarrow}

\usepackage{marginnote}

\begin{document}

\vspace*{1mm}

\noindent \makebox[11.5cm][l]{\small \hspace*{-.2cm} }{\small FERMILAB-PUB-19-147-T}  \\  [-1mm]

\begin{center}
{\Large \bf Luminous Signals of Inelastic Dark Matter \\[2mm] in Large Detectors}
\\[2mm]

\vspace{10mm}

{Joshua Eby$^1$, Patrick J. Fox$^2$, Roni Harnik$^2$, and Graham D. Kribs$^3$}

\vspace*{5mm}

\noindent 
$^1$Department of Particle Physics and Astrophysics, \\
    Weizmann Institute of Science, Rehovot 76100, Israel \\
$^2$Theoretical Physics Department, 
    Fermilab, 
    Batavia, IL 60510 USA \\
$^3$Department of Physics, 
    University of Oregon, 
    Eugene, OR 97403 USA 
\end{center}

\vspace*{-5mm}

\begin{abstract}
We study luminous dark matter signals in models with inelastic scattering.
Dark matter $\chi_1$ that scatters inelastically off elements in 
the Earth is kicked into an excited state $\chi_2$ that can subsequently 
decay into a monoenergetic photon inside a detector.  
The photon signal exhibits large sidereal-daily modulation due to the daily rotation of the Earth and anisotropies in the problem: the dark matter wind comes from the direction of Cygnus due to the Sun's motion relative to the galaxy, and the rock overburden is anisotropic, as is the dark matter scattering angle. 
This allows 
outstanding separation of signal from backgrounds.  
We investigate the sensitivity of two classes of large 
underground detectors to this modulating photon line signal:
large liquid scintillator neutrino experiments, including Borexino 
and JUNO, and the proposed large gaseous scintillator directional detection 
experiment CYGNUS\@.  
Borexino's (JUNO's) sensitivity exceeds the bounds from xenon experiments on inelastic nuclear recoil for mass splittings $\delta \gtap 240 (180)$~keV, and is the \emph{only} probe of inelastic dark matter for ${350 \text{ keV} \ltap \delta \ltap 600 \text{ keV}}$\@.
CYGNUS's sensitivity is at least comparable to
xenon experiments with $\sim 10 \; {\rm m}^3$ volume
detector for $\delta \ltap 150$~keV, 
and could be substantially better with larger volumes
and improved background rejection.
Such improvements lead to the unusual situation  
that the inelastic signal becomes the superior way to 
search for dark matter even if the elastic and inelastic
scattering cross sections are comparable.
\end{abstract}

\newpage

\section{Introduction}

Inelastic dark matter \cite{Hall:1997ah,TuckerSmith:2001hy,TuckerSmith:2004jv,Finkbeiner:2007kk,Arina:2007tm,Chang:2008gd,Cui:2009xq,Fox:2010bu,Lin:2010sb,An:2011uq,Pospelov:2013nea,Dienes:2014via,Barello:2014uda,Bramante:2016rdh} --
where dark matter scatters off nuclei into 
an excited state -- 
provides interesting new signals of
dark matter.  While originally motivated by the DAMA/LIBRA annual modulation \cite{TuckerSmith:2001hy,TuckerSmith:2004jv,Chang:2008gd,Cheung:2009qd,Graham:2010ca,Lin:2010sb,Barello:2014uda}, an inelastic dark matter explanation 
is now extremely difficult to reconcile with current data from 
several different experiments, despite valiant model building efforts \cite{ArkaniHamed:2008qn,Alves:2009nf,Lisanti:2009am,Chang:2010en,Schwetz:2011xm,Weiner:2012gm,Barello:2014uda}. 
One of the appeals of inelastic dark matter is that it 
provides a rationale for why dark matter has not yet been seen
in direct detection experiments.  Models of inelastic dark matter
typically involve an inelastic scattering cross section off nuclei that is much larger than the elastic cross section.  
Once the inelastic splitting is large, 
say $\delta \gtap 300$~keV, ordinary $Z$-exchange can mediate
$\chi_1 + N \ra \chi_2 + N$, and not be in violation of existing 
direct detection experimental bounds \cite{Bramante:2016rdh}. 
To probe large inelastic splittings it is critical to analyze
high recoil events \cite{Bramante:2016rdh}.  XENON100 \cite{Aprile:2017aas} 
and PandaX \cite{Chen:2017cqc} are already probing the inelastic frontier; 
however their sensitivity is rapidly diminished above about 
$300$~keV due to the intrinsic limitation that the heaviest element 
on which the dark matter can inelastically scatter is xenon.  
This motivates new ideas to probe inelastic dark matter.

In this paper, we demonstrate that large underground neutrino 
detectors as well as large directional dark matter detectors can provide 
a new method to probe inelastic dark matter.  
Other proposed searches for dark matter that could yield a signal in 
neutrino detectors include dark matter that destroys target baryons \cite{Davoudiasl:2010am,Davoudiasl:2011fj,An:2012bs,Huang:2013xfa}, 
dark matter that yields annihilation or decay products detectable 
in these experiments \cite{Yuksel:2007ac,Agashe:2014yua,Berger:2014sqa,Kong:2014mia,Alhazmi:2016qcs,Cui:2017ytb,Kachulis:2017nci,Campo:2017nwh,Kim:2018veo,McKeen:2018pbb,Bramante:2018tos},
self-destructing dark matter \cite{Grossman:2017qzw}, dark matter produced 
at high-intensity accelerators or radioactive sources \cite{Bringmann:2018cvk}, 
or dark matter bounced off energetic cosmic rays \cite{Bringmann:2018cvk,Ema:2018bih}.

\begin{figure}[t]
    \centering
    \includegraphics[width=\textwidth]{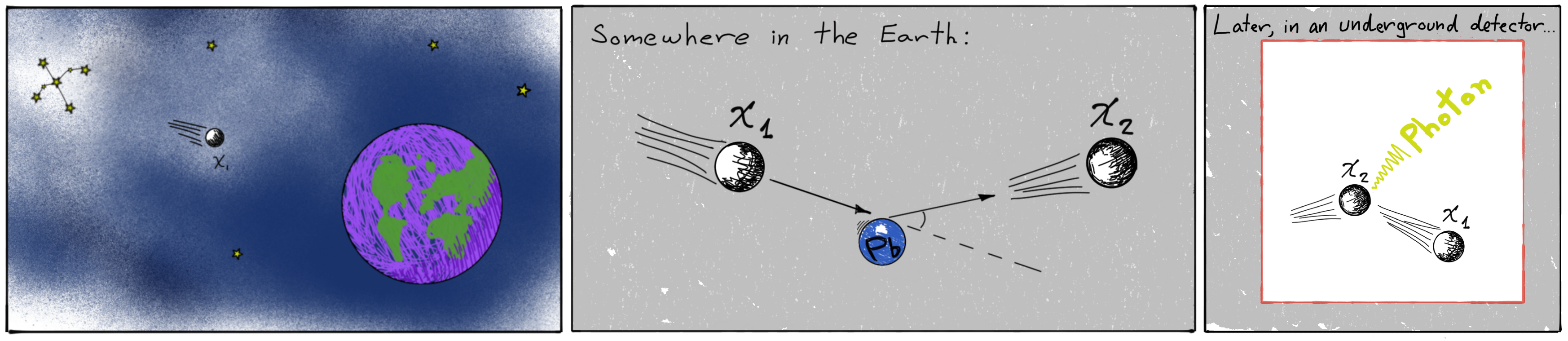}
    \caption{A cartoon summarizing the luminous  dark matter signal discussed in this work.  A heavy dark matter particle, $\chi_1$, is coming from the direction of the Cygnus constellation and approaching Earth with high speed (left). This allows it to upscatter off a (lead) nucleus somewhere within the Earth, deviating from its direction only slightly (middle). The excited state, $\chi_2$, decays back to $\chi_1$ and a photon in an underground detector located on the opposite side of the Earth (right). The rate for this process would be much lower had the detector been on the Cygnus-facing side of the planet.}
    \label{fig:cartoon}
\end{figure}

Consider a dark matter 
particle with mass of order a TeV traversing through the Earth.  This 
particle moves with the dark matter wind that appears in the Earth's frame 
to be coming from the Cygnus constellation. If its speed is
large enough, the dark matter can upscatter off a heavy nucleus (perhaps lead), into an excited state which could be several hundreds of 
keV heavier in mass.  Had this upscattering been attempted off a 
xenon nucleus in one of the large direct detection experiments, 
it would have been kinematically inaccessible. The excited state continues 
in the same general direction, headed from Cygnus, for tens to hundreds 
of kilometers before decaying inside a large and clean detector, such as Borexino, 
JUNO, or the planned CYGNUS dark matter detectors. This type of signal exhibits 
a strong daily modulation because, when the detector is on the Cygnus-facing side 
of the Earth, there is far less ``target material'' available to upscatter. 
The mechanism is summarized, in cartoon form, in \figref{fig:cartoon}.

The signal we will consider thus consists of single photons with an energy equal to the mass splitting $\delta$ of inelastic dark matter and with a rate that exhibits a strong modulation with a period of a sidereal day. The shape and phase of the modulation is predicted and depends on the geographical location of the detector, allowing for excellent signal to background discrimination. 
The essential ingredients are: inelastic dark matter which is more massive than the target nucleus and mass splitting anywhere from the threshold of the detection experiment up to about $600$~keV\@. The lifetime, ideally in the range of $0.1$-$10$ seconds, is also crucial, but can arise naturally through a radiative decay.

Similar ingredients have been considered before in 
other contexts: 
Luminous Dark Matter \cite{Feldstein:2010su} 
first proposed dark matter which upscatters to an excited state in the rock 
outside of a detector, and then decays into a photon that needed 
to be a few keV, with the goal of explaining the DAMA annual modulation.  
Alas, this novel DAMA explanation is ruled out by other direct detection experiments. 
Another idea, ``dark matter in two easy steps'' \cite{Pospelov:2013nea},
proposed dark matter which upscatters into an excited state in lead 
shielding surrounding a neutrinoless double beta decay experiment, and 
then decays into a photon of order one hundred keV\@.  This analysis
shares some similarities with our paper: we both propose
that inelastic dark matter is excited by lead outside of the 
detector volume, and propose looking for the photon from the decay
of the excited state back into dark matter.  Moreover, we both
recognize that there is large sidereal-daily modulation 
of the rate that can be used to separate signal from background.
The main differences between our analysis and theirs is: 
we utilize the entire Earth as upscatter material (focusing on 
upscatters off lead and iron); and, we consider a large range of 
photon energies between about $5$-$600$~keV\@.  The upper end of this
range has the weakest constraints from direct detection experiments.
The range $75$-$125$~keV, which was the main focus of
\cite{Pospelov:2013nea}, 
is significantly constrained by 
PICO and the high recoil analysis of XENON100 \cite{Bramante:2016rdh}.
Reinterpreting these bounds on the magnetic inelastic transition
strength \cite{Bramante:2016rdh} implies the 
the characteristic decay length of the excited state exceeds 
50 (500) meters once $\delta \ltap 150 \,(100)$~keV\@.  
At these lengths, the gain from integrating over the trace lead
abundance in the Earth is substantial in comparison to utilizing the 
close-in concentrated lead shielding of experiments sensitive to
these photon energies. 

In this paper we also decouple the upscatter process from the excited
state decay.  This is well-motivated, since 
there are many specific models that can have a large inelastic 
scattering rate, small elastic scattering rate, and excited
state decay that all proceed through different processes.
Examples include a variety of elementary candidates such as the wino and higgsino 
\cite{Hisano:2010fy,Hisano:2011cs,Hill:2013hoa,Hill:2014yka}
as well as composite candidates \cite{Kribs:2016cew}.
The prototypical example model for this paper is the 
narrowly-split higgsino that arises from split
Dirac supersymmetry \cite{Fox:2014moa}:  in this model,
the abundance of dark matter matches cosmological observations
for $m_{\chi^0_1} \sim 1$~TeV; 
the neutral higgsino states $\chi^0_{1,2}$ are narrowly split 
by hundreds of keV; the dominant elastic scattering cross section
is exceptionally small due to the twist-2 operator suppression
as well as the cancellation against Higgs boson 
exchange \cite{Hill:2013hoa,Hill:2014yka};
and, the dominant decay of the excited (neutral) higgsino
is indeed $\chi^0_2 \ra \chi^0_1\, \gamma$ \cite{Haber:1988px}. 
But, we emphasize that none of these characteristics are unique
to higgsinos -- other models could easily yield similar outcomes.

What kind of detectors have the best sensitivity to photon line emission arising from the decay $\chi_2^0 \ra \chi_1^0\, \gamma$?
Dark matter detectors can certainly be sensitive
to this signal, and in fact they are designed to observe much 
smaller energy depositions.  But, these direct detection
experiments are smaller in size relative to their neutrino
and directional dark matter detector counterparts, and they also 
have significant backgrounds when considering that there is only 
an electron-equivalent energy deposition from the photon.  
At xenon detectors, 
the background rate before cutting on a fiducial region is 
roughly at level of $0.1$~events/kg/day/keVee
\cite{Aprile:2011vb,Akerib:2014rda}. 
Within the fiducial region, the irreducible background 
from radioactive impurities (radon and krypton) is
roughly $10^{-3}$~events/kg/day/keVee \cite{Aprile:2017fhu}.
Borexino has a scintillator mass of $278$~tons
with a ``background'' rate of 
roughly $10^{-6}$~events/kg/day/keV \cite{Bellini:2013lnn}.
For our purposes in this paper, background means the genuine 
radioactive background combined with the solar neutrino 
``background'' (satisfying the common idiom that yesterday's 
signal is today's background).  This low background rate does, 
however, rise rapidly for energy depositions below about 
$250$~keV, where the signal must compete against a large 
background rate from $^{14}$C $\beta$-decay, see \figref{fig:backgrounds}.

It is clear that the particular material \emph{within} the detector 
is not important so long as there is a high efficiency to absorb 
the photon from the excited state decay 
in order to generate a scintillation signal.
Indeed, what maximizes our signal is the instrumented 
\emph{volume} of a detector.  This suggests
large volume directional dark matter
detectors which utilize a gaseous scintillator would be ideally suited
to maximize the sensitivity to the photon deposition, if the 
backgrounds can be minimized.  Certainly one obvious advantage of 
a gaseous scintillator is that its much-reduced density automatically
reduces backgrounds from scattering, such as from solar neutrinos.
Below we also discuss the opportunities for the proposed CYGNUS detector 
to probe the photon signal from inelastic dark matter.

\section{Illuminating Inelastic Dark Matter}

Inelastic dark matter is characterized by dark matter scattering
off nuclei predominantly into an excited state with an  
elastic scattering rate below existing bounds.  
The range of inelastic splittings allowed by direct detection
experiments that probed the nuclear recoil signal resulting from
an inelastic collision was carried out in Ref.~\cite{Bramante:2016rdh}.  
In this paper, our focus is on the detection of the excited state decay.  

\subsection{Model-independent requirements}

Three basic requirements are needed to have a possibility of experimentally detecting 
excited state decay into dark matter: 
\begin{enumerate}
\item  Inelastic scattering $\chi_1 N \ra \chi_2 N$ off (typically heavier) 
elements in the Earth has a sufficiently  
large cross section to populate the flux of the excited state passing through 
a suitable detector.  
\item  The excited state decays frequently via $\chi_2 \ra \chi_1 \gamma$, 
with a lifetime that is long compared with the transit time through 
the detector, but not substantially longer than the transit time through the Earth.
\item  Ordinary dark matter direct detection experimental bounds are 
satisfied. This constrains both (\emph{a}) spin-independent elastic scattering, and (\emph{b}) 
inelastic scattering off nuclei \emph{within} a direct detection experiment.
\end{enumerate}
In addition, if dark matter is heavier than the target nucleus, the upscattering
yields an excited state that continues mainly in the forward direction,
which results in a signal with a significant sidereal-daily modulation.

The requirement that the spin-independent elastic scattering be
below current experimental bounds, item~3(\emph{a}), is satisfied by a wide range of models.  We have already mentioned the higgsino and wino as examples
where the spin-independent elastic scattering cross section is very small
\cite{Hill:2013hoa,Hill:2014yka}. Composite dark matter 
candidates in this mass range can also have strong suppression if the
leading spin-independent scattering interaction is an effective operator 
of high dimension, for example stealth dark matter that proceeds through
the dimension-7 polarizability operator \cite{Appelquist:2015zfa}.

The model space is narrowed by the need to fulfill both requirement 1 and 3(\emph{b}), namely that inelastic scattering
off nuclei in the Earth has a sufficiently large cross section 
which, nevertheless, is not itself ruled out by inelastic scattering 
off the elements within a direct detection experiment.  When
$\delta \gtap 250$~keV, the bounds from direct detection 
experiments become highly suppressed by the very small fraction
of the dark matter velocity distribution that can scatter
(as well as the suppression from the nuclear form factor).
In this region, inelastic dark matter can scatter much more easily
off the heaviest trace elements in the Earth, and this provides
a major lever-arm against direct detection bounds. 

A second possibility, especially for smaller $\delta \ltap 150$~keV, 
is that the inelastic cross section is itself very small.  The 
advantage of considering such small $\delta$ is that there are
lighter elements in the Earth, such as iron and silicon, 
that have a much larger number density.  Here, a large underground 
experiment could have superior sensitivity over  
present direct detection experiments simply due to 
its (much) larger volume.  

Finally, the requirement that the decay $\chi_2 \ra \chi_1 \gamma$ occurs frequently 
is not especially onerous.  Kinematically there are only two possibilities 
in the Standard Model:  the 2-body photon decay and the 3-body 
process $\chi_2 \ra \chi_1 \nu \bar{\nu}$.  Phase space favors 
the photon decay, though a model-dependent calculation is needed 
to determine which process actually dominates. We shall see that in an interesting class of models this is indeed the case and that the lifetime is also in the interesting range.  We note that if a light mediator with mass less than $\delta$ is also present in the 
model, then additional decay modes can be present.

\subsection{Narrowly-split higgsinos}
\label{sec:genxmodelbuildingtotherescue}

The reader may find it helpful to have a concrete model in 
mind while considering our model-independent results below. For this,
the narrowly-split higgsino, which we describe below, 
provides a great example. 
For a recent review of the experimental
status of the narrowly-split Higgsino, see~\cite{Krall:2017xij}.
Readers who have no appetite for models in these data-driven times may move on to the next section. 

Higgsinos are spin-$1/2$ superpartners of the two Higgs doublets in the
minimal supersymmetric standard model \cite{Martin:1997ns}. In the limit
that all of the superpartners are heavy, the higgsino spectrum is characterized by
\begin{eqnarray}
m_{\chi} &\simeq& |\mu|  \nonumber \\
m_{\chi^\pm} - m_{\chi^0} &\simeq&  \left( 1+\frac{1}{\cos\theta_W} \right) \alpha_2 \,m_W\, \sin^2 \left(\frac{\theta_W}{2}\right) \\
m_{\chi^0_2} - m_{\chi^0_1} &\simeq& m_Z^2 
  \left( \frac{\sin^2\theta_W}{M_1} + \frac{\cos^2\theta_W}{M_2} \right) 
  \nonumber
\end{eqnarray}
where $\mu$ is the higgsino mass and $M_{1,2} \gg |\mu|$ are the Majorana masses for 
the bino and wino \cite{Cirelli:2005uq,Fox:2014moa}.  It is well known that the 
higgsino dark matter 
relic abundance matches cosmology when $|\mu| \simeq 1.1$~TeV\@.  This mass scale is
obviously above the mass of the heaviest nucleus we consider in this paper (lead), 
consistent with maximizing the inelastic scattering rate. A narrow splitting 
$\delta = m_{\chi_2^{}} - m_{\chi_1^{}} \ltap 800 \; {\rm keV}$
between the neutral higgsinos occurs when $M_{1,2} \gtap 10^7 \; {\rm GeV}$. 
The large splitting between the higgsino mass and the electroweak gauginos
could arise naturally as a one-loop radiative correction from a heavy bino  \cite{Fox:2014moa}. 
The elastic scattering cross section is highly suppressed, 
$\sigma_{\chi_1^{} n} \ltap 10^{-48} \; {\rm cm}^2$,
due to the twist-2 operator 
suppression as well as the partial cancellation against Higgs boson 
exchange \cite{Hill:2013hoa,Hill:2014yka}.
The excited state neutral higgsino has a one-loop radiative decay 
with width (in the limit where all other superpartners are 
decoupled \cite{Haber:1988px})
\begin{eqnarray}
\Gamma_{\chi_2^{} \rightarrow \chi_1^{}\gamma} &\simeq& 
  \alpha_{\rm em}\, \alpha^2_W\, \frac{\delta^3}{4\pi^2 m_{1}^2}
  \label{eq:decaywidth}
\end{eqnarray}
that leads to a characteristic decay length of
\begin{eqnarray}
\ell_{\chi_2^{}} = \frac{c v}{\Gamma_{\chi_2^{} \rightarrow \chi_1^{} \gamma}} &=& 20 \; {\rm km} \left( \frac{c v}{400 \; {\rm km/s}} \right) \left( \frac{400 \; {\rm keV}}{\delta} \right)^3 \left( \frac{m_{1}}{1 \; {\rm TeV}} \right)^2 \, ,
\label{eq:decaylength}
\end{eqnarray}
where $m_{(1,2)} = m_{\chi^{}_{(1,2)}}$.
The photon decay can be compared with the 3-body process
mediated by the weak interaction
\begin{eqnarray}
\Gamma_{\chi_2^{} \rightarrow \chi_1^{} \nu\bar{\nu}}
&\sim& \alpha^2_W \frac{\delta^5}{120 \cos^4\theta_W \pi m_Z^4} \, .
\end{eqnarray}
The photon decay dominates provided
\begin{eqnarray}
\left( \frac{\delta}{1 \; {\rm GeV}} \right)^2 \left( \frac{m_{1}}{1 \; {\rm TeV}} \right)^2 &\ltap& 1 \, . 
\end{eqnarray}
This is clearly satisfied throughout the inelastic parameter space that
we consider in this paper.

Finally, the inelastic transition $\chi_1^{} N \ra \chi_2^{} N$ is mediated
by $Z$-exchange for the higgsino that (famously) has a per nucleon 
cross section 
$\sigma_0\sim 10^{-39}$~cm$^2$, before kinematic suppression from phase space and
nuclear form factors are taken into account.  Once $\delta \gtap 300$~keV,
direct detection experimental bounds are satisfied 
 \cite{Kribs:2016cew,Aprile:2017aas,Chen:2017cqc}. 
In the narrowly-split higgsino model, the inelastic transition proceeds 
through $Z$-exchange, and so therefore the inelastic cross section is fixed.  
If we generalize beyond $Z$-exchange, or allow $|\mu|$ to take values 
somewhat smaller or larger than the relic abundance would suggest, 
there is larger range of the inelastic scattering cross section 
that could be considered. As we will see,
the detection of photons in large underground neutrino experiments will
extend the reach for inelastic dark matter to both larger inelastic splittings
as well as cross sections that are considerably smaller than $Z$-exchange.

For the remainder of the paper, we assume a model-independent spin-independent inelastic scattering cross section $\sigma_0$ (not necessarily $Z$-exchange) with an excited state 
decay rate given by \eref{eq:decaywidth}.

\section{Luminous Signals and Sidereal Daily Modulation}
\label{sec:luminous}

In this section we discuss the luminous dark matter signals in inelastic models and show that the signal rate modulates strongly with a period of one sidereal day. The modulation is the result of a convolution of three anisotropic effects, which will be discussed in the upcoming three subsections: the anisotropy of high-speed dark matter, the anisotropy of the rock overburden in the lab, and the anisotropy of the upscattering of heavy dark matter in the Earth. 

\subsection{Anisotropy of high speed dark matter} \label{sec:anisotropyspeed}

It is well known \cite{McCabe:2013kea} that the motion of the 
Sun/Earth system in the galaxy causes  
the dark matter wind to head towards us from a direction that
approximately aligns with the location of the Cygnus constellation. 
Yet, dark matter impinging the Earth is not unidirectional, 
since the velocity distribution of dark matter in the galactic frame 
extends up to the escape velocity $\simeq 550$~km/s. This is more 
than a factor of 2 larger than our speed relative to this frame,
$v_{\rm sun} \simeq 220$~km/s.
Inelastic dark matter, however, can scatter only once the 
dark matter speed is high enough in the lab frame to overcome
the inelastic transition.  For large inelastic transitions
($\delta \sim 300$-$600$~keV), these high speeds can be obtained 
only by combining the high velocity tail of the 
galactic frame dark matter distribution with the boost into 
the Earth frame.  Or, in other words, the very highest speeds of 
dark matter perceived in the Earth frame are unidirectional 
from the Cygnus constellation.

\begin{figure}[!h]
\begin{center}
 \includegraphics[width=0.40\textwidth]{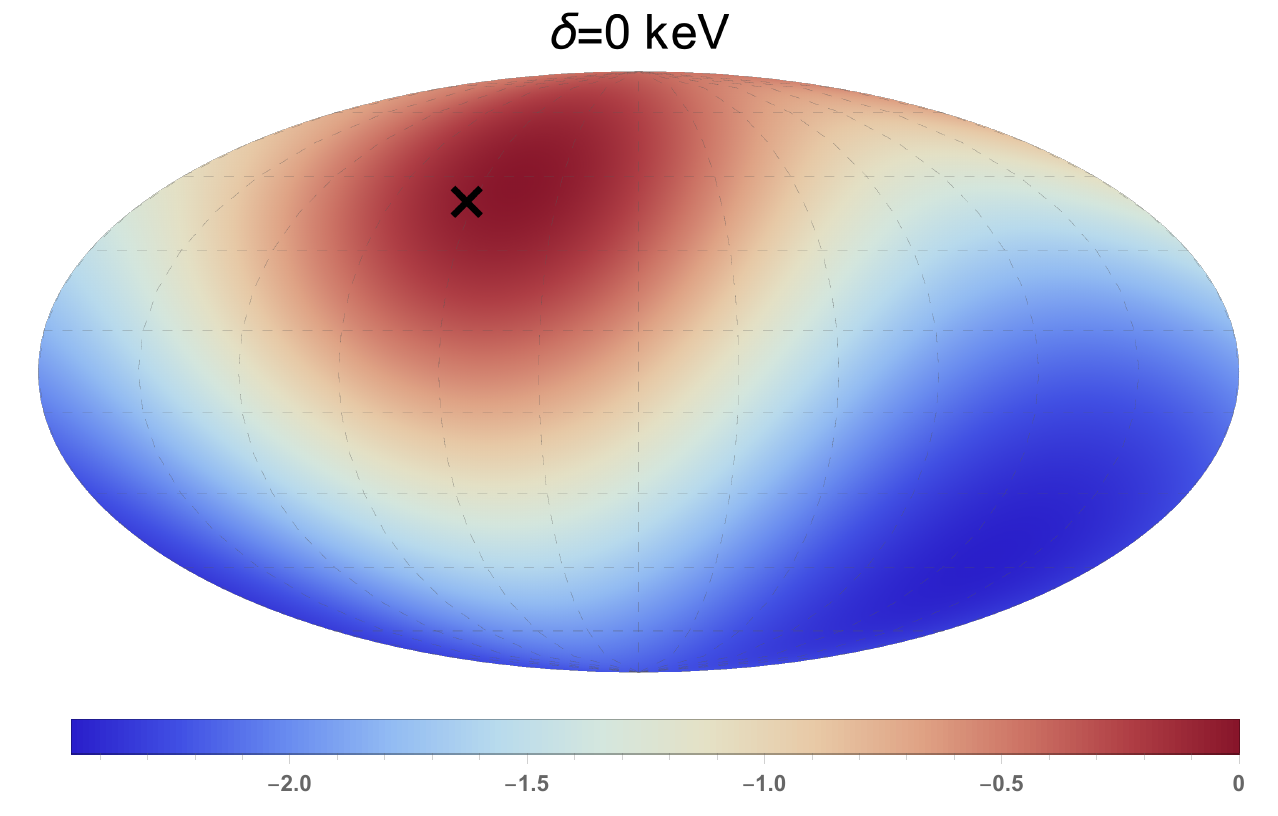}
 \hspace*{0.05\textwidth}
 \includegraphics[width=0.40\textwidth]{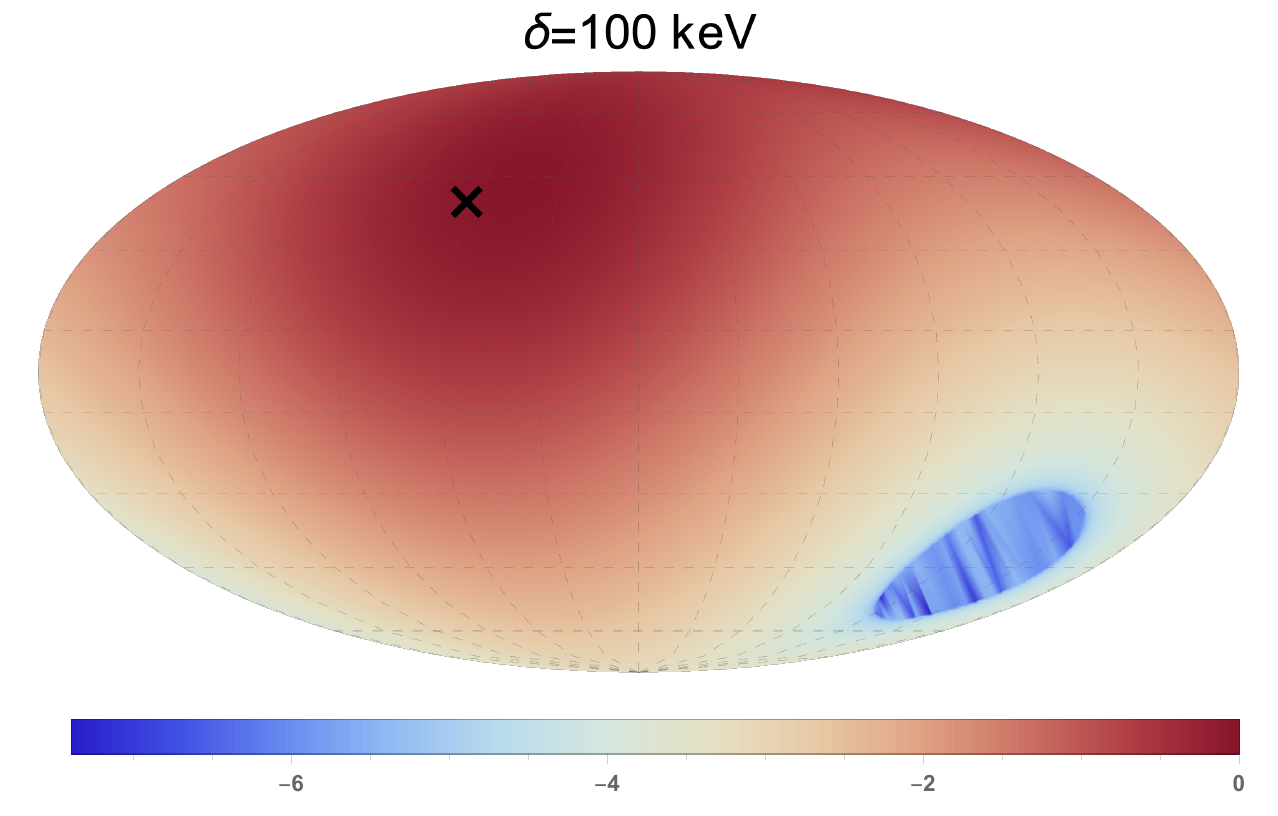}\\
 \includegraphics[width=0.40\textwidth]{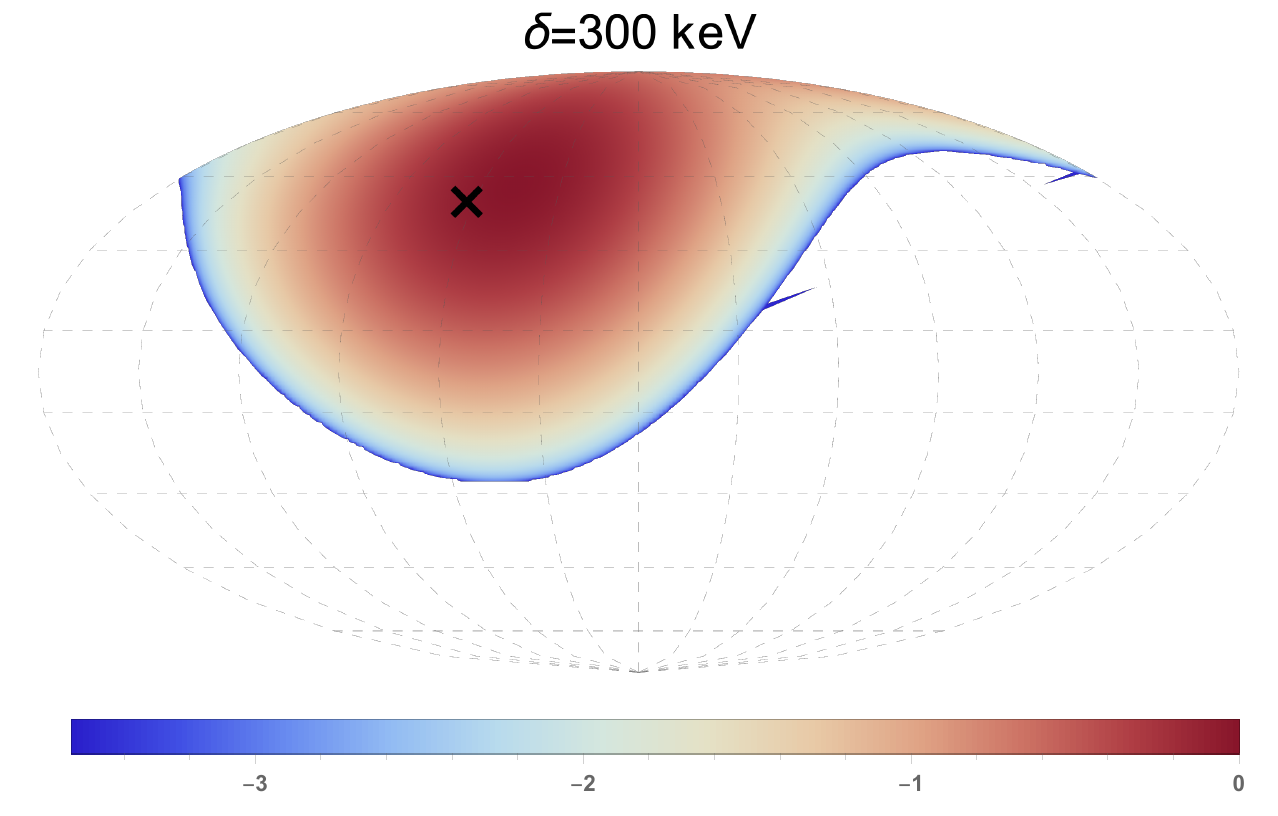}
 \hspace*{0.05\textwidth}
 \includegraphics[width=0.40\textwidth]{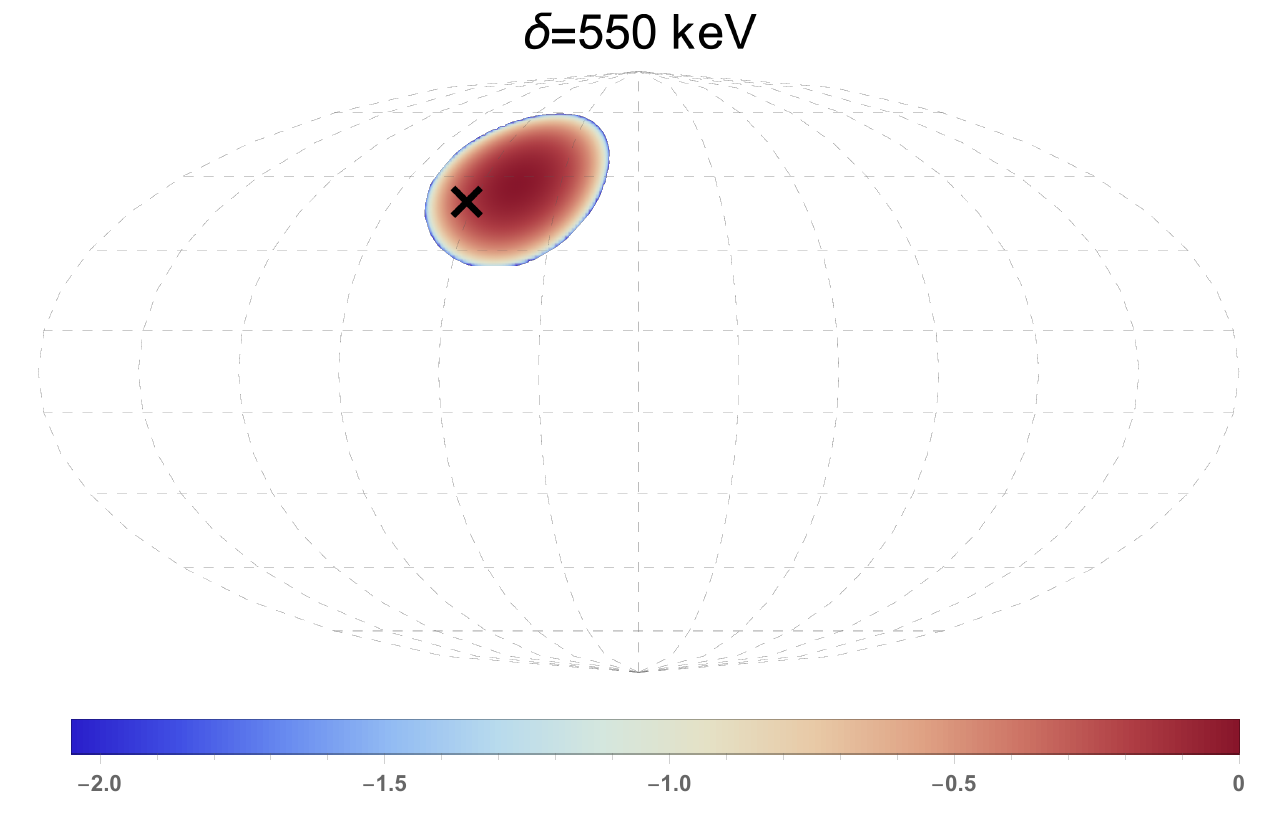}\\
 \includegraphics[width=0.8\textwidth]{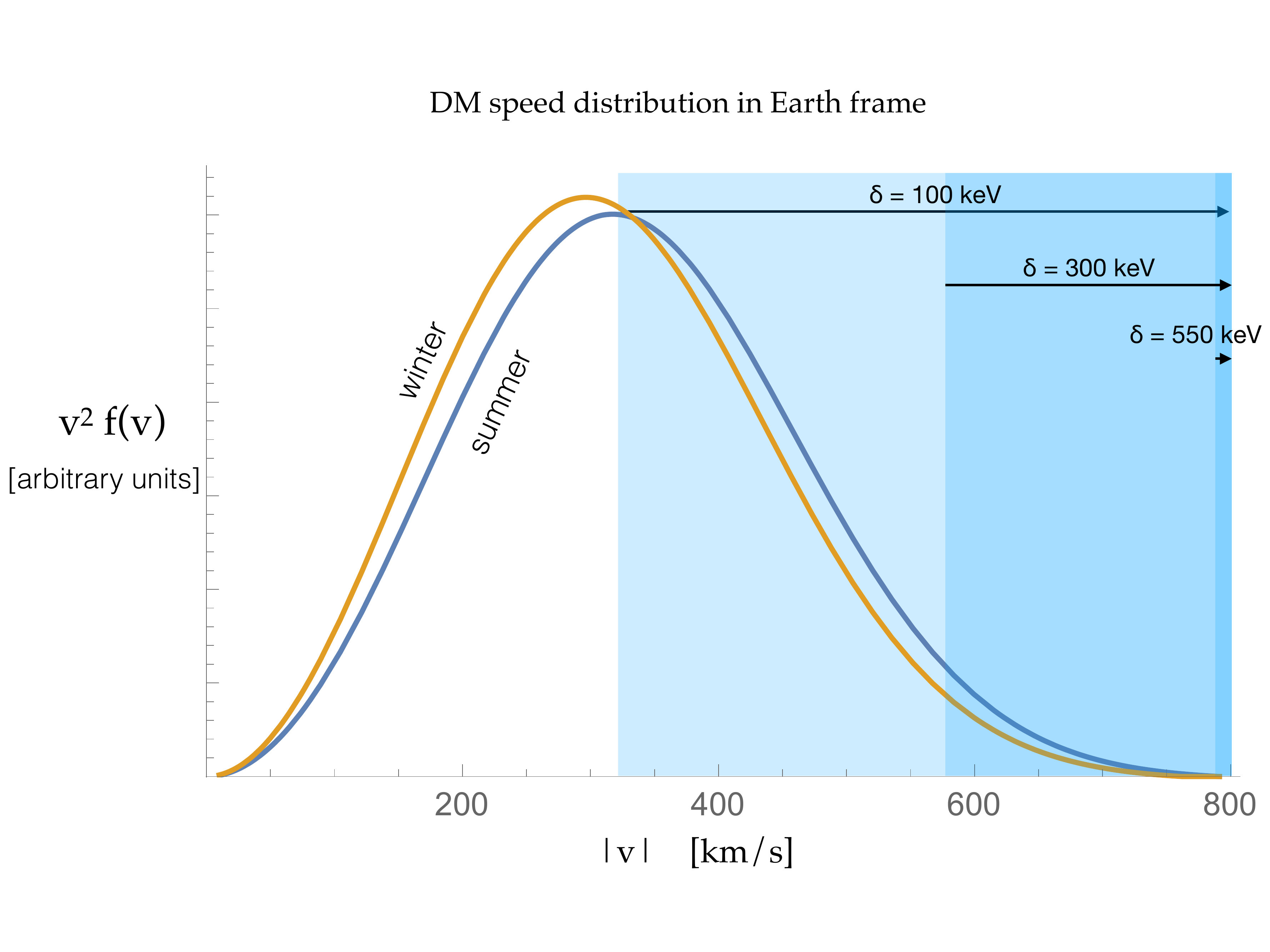}
 \vspace*{-3em}
\end{center}
\caption{Upper four panels:  The flux of 1 TeV dark matter capable of scattering off lead and its dependence on inelastic splitting, $\delta$.  The colors denote $\log_{10}$ of the ratio of the flux to the peak flux, which comes from the direction of motion of the Earth, which is given by the cross and is approximately in the direction of the Cygnus constellation.
Lower panel: The fraction of the velocity distribution of dark matter in the
Earth frame that can scatter off lead for the same values of $\delta$.}
 \label{fig:fluxCygnus}
\end{figure}

As $\delta$ is decreased from the maximum that permits any
inelastic scattering, 
the range of speeds of dark matter which can scatter becomes larger.
This also broadens the range of arrival directions from which dark matter can still 
inelastically scatter.
We illustrate this in \figref{fig:fluxCygnus}. 
The dark matter flux, 
when viewed as a distribution projected onto the sky in Earth frame (top panels), 
decreases rapidly away from the peak at Cygnus. 
Dark matter appearing from an increasingly larger region 
around Cygnus necessarily samples the increasingly 
suppressed dark matter velocity distribution tail.  

Unlike most other terrestrial beings that are interested in the position of the Sun,
a dark matter direct detection experimenter ought to be interested in the position of Cygnus.
Most dark matter particles, particularly those with high speed, appear to originate from the vicinity of Cygnus. This is obviously of importance if the detector has directional sensitivity \cite{Mayet:2016zxu}.  Another example are models in which dark matter is sufficiently strongly interacting to have a high probability to scatter in the overburden. 
The rate in a detector will then depend on 
whether the line from the detector to Cygnus goes through $\sim 1$~km of rock above typical underground detectors or the entire diameter of the Earth (see e.g.~\cite{Emken:2017qmp}).    
In most ``usual'' weakly-coupled dark matter scenarios, by contrast, scattering off 
elements in the Earth occurs so rarely that the direction to Cygnus is irrelevant. 

In our scenario, the overburden of the Earth is \emph{critical} to obtaining a high signal rate.
Dark matter inelastically scatters off material in the Earth, emerging as the excited state. As~\figref{fig:fluxCygnus} shows, this implies the dark matter is most likely coming from the direction of Cygnus. 
During the hours that Cygnus is below the horizon the scattering may be, for example, in the Earth's crust or mantle.
The excited state travels a distance on average between
$\sim 10$--$1000$~km, then decays into the dark matter plus 
a photon with energy $E_\gamma = \delta$.  When these decays
occur inside an instrumented region, the photon can be detected.
When Cygnus is above the horizon, there is a much smaller amount
of overburden off which the dark matter can inelastically scatter,
and a correspondingly smaller probability for the excited state
to decay inside the detector since the detector depth is much
smaller than the decay length.  
It is interesting that this is distinct from models with a very large scattering cross section, for which a large overburden leads to a substantially reduced rate~\cite{Emken:2017qmp}. 
In either case, the orientation of the detector relative to Cygnus is of paramount importance.

While the Sun returns to the same position in the sky every 24 hours -- a solar day -- the fixed stars return every \emph{sidereal} day which is approximately 4 minutes shorter.  Ignoring the very small shift (parallax) in the location of stars due to the Earth's rotation around the Sun\footnote{As well as the eccentricity of the Earth's wobble that occurs on a much longer timescale than is relevant for dark matter direct detection.}, all of the fixed stars (and constellations) rise and set at fixed sidereal times that are determined solely by the declination of the star in the Earth's reference frame and the location on Earth.

Cygnus has a declination of approximately 45$^\circ$ North.
This means detectors in the Northern Hemisphere at
latitudes above 45$^\circ$ North see Cygnus 
above the horizon at all times of the sidereal day.  
Conversely, (hypothetical) detectors in the Southern 
Hemisphere below 45$^\circ$ South never see Cygnus above the 
horizon. In this paper we consider primarily three locations for detectors:
Gran Sasso at 42.6$^\circ$ North (home to the Borexino experiment;
also one of the locations for the CYGNUS directional dark matter experiment),
Jiangmen at 22.1$^\circ$ North (where the JUNO experiment will be located), 
and SUPL at 37.1$^\circ$ South (where we consider a hypothetical 
Borexino-like experiment for the purposes of studying the unusual
signals that could be seen in the Southern Hemisphere).  
In all of these cases, 
Cygnus is below the horizon for part of each sidereal day --
about 2.5 hours for Gran Sasso, 8.5 hours for Jiangmen, and 18.5 hours
for SUPL\@.  
The modulation of our signal will be determined by this sidereal schedule or the rise and set of Cygnus.

\subsection{Anisotropy of the overburden}

Since our signal depends on the overburden which dark matter must traverse to reach a detector, it is worthwhile studying the overburden, including its anisotropy and elemental composition.

\begin{figure}[t]
\begin{center}
\includegraphics[width=0.49\textwidth]{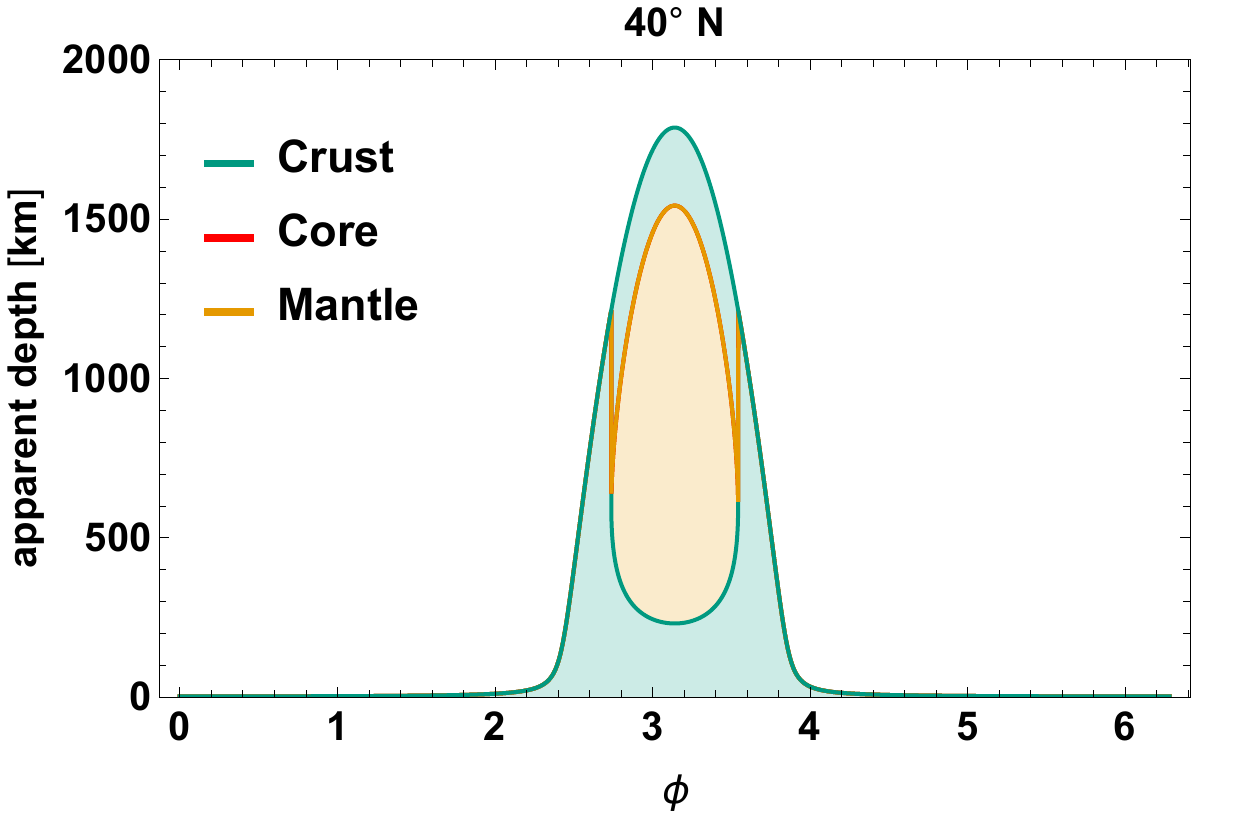}
\includegraphics[width=0.49\textwidth]{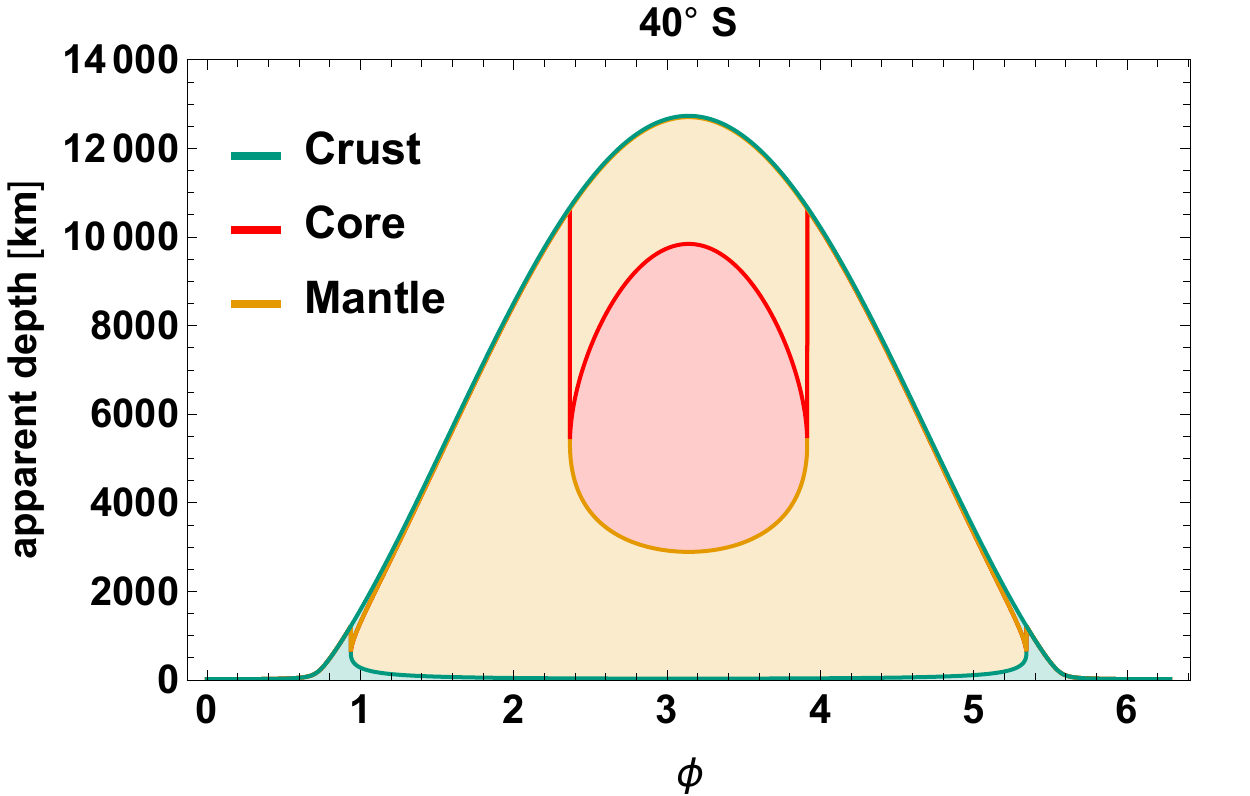}
 \caption{The distance through the Earth from a point $2$ km below the Earth's surface at $40^\circ$ North (left) and South (right) of the equator, along the direction to Cygnus, through a (sidereal) day; $0<\phi\le 2\pi$.}
 \label{fig:apparentdepths}
 \end{center}
\end{figure}

We can gain some insight into daily modulation effects by considering the 
apparent depth of a lab along the direction to Cygnus.  The center of the Cygnus constellation 
is at about $45^\circ$ declination, which is close to the 
latitude of Gran Sasso, see \tabref{tab:labs}, meaning that the apparent 
depth of Gran Sasso is never particularly large.  However, for a lab whose latitude lies below the declination of Cygnus, the depth becomes larger.  We show in \figref{fig:apparentdepths} 
the apparent depth for two underground labs, each at a depth of $2$ km, 
at $40^\circ$ North and South of the equator.  

How much of the apparent depth is relevant to our signal depends on the 
lifetime of the excited state.  As we saw from 
\eref{eq:decaylength}, typical decay lengths range from 
$\sim 10$--$1000$~km for $\delta \sim 550$--$100$~keV respectively,
for dark matter with a mass of $m_1 = 1$~TeV\@.  These lengths are 
fascinating, because they are significantly larger than 
the depth of underground detectors, 
but they may be considerably shorter than the Earth's radius. 
As a result the signal rate depends on the composition of the Earth.

For heavier dark matter, $m_{\chi_1^{}} \gtap m_{\rm nucleus}$,
and large inelastic splittings $\delta \gtap 150$~keV, 
the element with the largest probability to inelastically 
scatter off is lead.  Lead is the most massive
element in the Earth which is both relatively abundant and stable. 
Its abundance is roughly $10^{-5}$ g/g in the Earth's crust \cite{crust} and a factor of five smaller in the mantle~\cite{mantle}.
Moreover, lead (with $A \simeq 210$) is heavier than all of the elements 
used in direct detection experiments, providing a kinematic 
advantage for highly inelastic dark matter $\delta \gtap 350$~keV
where there is no bound from other experiments. For lighter dark matter and/or lower splittings we will also consider scattering off of iron and silicon.

The details of the signal may also depend on geology, particularly on the difference between the crust and the mantle.
In \figref{fig:apparentdepths} we show separately the 
line of sight depth to Cygnus through the core, crust and mantle.  
Which of these layers dominates the total rate will depend upon the lifetime of the excited state and the relative abundances of elements that dark matter can upscatter. The full analysis, below, takes all of this into account. 

 In most cases the effect of crust-mantle differences are small; however there are models in which it is important. A notable curious example is a model with $\delta \gtap 300$~keV and detectors in the southern hemisphere. 
 In this case, the characteristic decay length is less than $100$~km, and due to it's higher lead abundance, the crust dominates over the mantle. As can be seen in the right panel of~\figref{fig:apparentdepths}, the apparent 
depth of crust peaks at two separate angles $\phi \simeq \pi/3, 5\pi/3$, which will lead to a novel double-peaked modulation pattern which will be shown in detail in~\secref{sec:resultsmodulatingrates}.

\subsection{Anisotropic scattering}

The third anisotropic effect which was implicitly assumed in the previous subsections, but is required in order get a large modulation of our signal rate, is the angular distribution of the outgoing (excited-state) dark matter particles following a scattering off a nucleus in the Earth. In the center of mass frame, the scattering of dark matter off a nucleus is generically isotropic.  However, if the dark matter is significantly heavier than the target nucleus, the center of mass frame and the lab frame are not coincident.
As a result, for heavy dark matter the isotropic distribution is boosted and the dark matter scattering is mostly forward in the lab frame.

In the lab frame, the maximal deflection angle of the outgoing dark matter for the elastic dark matter case is
\begin{equation}
\label{eq:deflection}
\cos\theta_\mathrm{max}^\mathrm{lab} = \sqrt{1-\frac{m_T^2}{m_{\chi_1}^2}}~,
\end{equation}
for a target mass $m_T$ and a dark matter mass of $m_{\chi^{}_1}$. 
For dark matter of order a TeV, the dark matter is deflected at most $\sim 10^\circ$ off its original course, even if it scatters off a lead nucleus. We shall see in the next subsection that in the inelastic case this result is qualitatively unchanged, though \eref{eq:deflection} does receive a correction, see \eref{eq:costhetamaxiDM}. The small deflection in the direction of the incoming dark matter, coupled with the anisotropic effects of the previous subsections, leads to a daily modulating event rate for heavy dark matter. In this work we will mostly focus on two target nuclei, iron and lead, and thus daily modulation is present for dark matter masses of a few hundred GeV or above. In Ref.~\cite{Feldstein:2010su}, where dark matter is light in order to address DAMA, the daily modulation effect is not present for the dominant scattering off iron. It may be interesting to consider the sidereal-daily modulating signal for light dark matter scattering off lighter elements in the Earth or even off electrons, but this is left for future investigation.

\subsection{Calculation of the modulating event rate}

While useful intuition can be gained from studying the overburden as a function of lab location or assuming all dark matter comes from the direction of Cygnus (see 
\secref{sec:anisotropyspeed}), the full calculation must take into account the excited state's lifetime, the distribution of incoming dark matter velocities, the position of the detector, form factors, and the distribution of target elements within the Earth.  We now present the details of the full calculation, which must be done numerically.

For a photon to be observed at an underground detector, an incoming dark matter particle must enter the Earth with sufficient speed that it can scatter off a target nucleus in the Earth at position $\vec{r}_s$, creating an excited state.  This excited state must scatter through the appropriate angle, such that it then travels, with velocity $\vec{v}_f$, towards the detector, which is at $\vec{r}_D$, and it must decay between entering and leaving the detector.  We take the scattering and the decay to be isotropic in their center of mass frames.  The amount of available scattering material grows as $|\vec{r}_s-\vec{r}_D|^2$ but the probability to scatter towards the detector scales as $|\vec{r}_s-\vec{r}_D|^{-2}$. Thus, all scatter sites within a decay length of the detector are (approximately) equally important.  Once the separation becomes larger than $v_f \tau$, decays become important.  The interplay between scatter site, elemental abundances and lifetime is complicated.

The total rate is calculated by integrating over all possible scatter sites, $\vec{r}_s$, in the Earth. To model the distribution of target nuclei, we use a three-layer model of the Earth corresponding to the core, mantle, and crust. The number density $n_T(\vec{r}_s)$ of targets is a function of $\vec{r}_s$ insofar as the density is different in the core \cite{core}, mantle \cite{mantle}, and crust \cite{crustmathematica}. We consider the cases where the target atoms are iron nuclei, lead nuclei, and to a lesser extent silicon nuclei; see \tabref{tab:Density}.  As might be imagined, determining the abundance of elements within the Earth is a challenging endeavor.  By comparison to the chemical composition of chondritic meteorites, rocks from the upper mantle and core samples of the crust, one can infer an abundance of each element in the core, mantle and crust.  The uncertainties on these numbers are at best $10-20\%$.  Furthermore, these are average abundances and there are undoubtedly large local variations.  We use only the central average value and do not attempt to incorporate uncertainties in abundances in our rate calculations.

 \begin{table}[t]
 \begin{center}
 \begin{tabular}{ c  c c  c  c }
 \hline \hline
  - & $n_{Si}$ [km$^{-3}$] & $n_{Fe}$ [km$^{-3}$] & $n_{Pb}$ [km$^{-3}$] & Outer Radius [km] \\
  \hline
  Core & $1.4\times 10^{37}$ & $1.0\times10^{38}$ & $1.3\times10^{31}$ & $3483$ \\
  Mantle & $2.1\times 10^{37}$ & $3.1\times10^{36}$ & $2.4\times10^{30}$ & $6341$ \\
  Crust & $1.7\times 10^{37}$ & $2.0\times10^{36}$ & $8.4\times10^{31}$ & $6371$ \\
  \hline \hline
\end{tabular}
\caption{Number densities for silicon, iron, and 
lead
, along with the outer-edge radii
, for the core \cite{core}, mantle \cite{mantle}, and crust \cite{crust,crustmathematica} in our three-layer approximation of the Earth.}
\label{tab:Density}
\end{center}
 \end{table}

\sloppy As discussed earlier (see \secref{sec:anisotropyspeed}), for an upscatter to occur in the collision, the incoming dark matter must have high speed in the lab frame, which leads to a strong directionality in flux.  In the galactic frame, the dark matter speed follows a Maxwell-Boltzmann distribution $f_{\mathrm{gal}}(\vec{v}_{\mathrm{MB}}) \propto \exp\left(-v_{\mathrm{MB}}^2/v_0^2\right) \theta(v_{\mathrm{esc}}-v_{\mathrm{MB}})$, and we take $v_0=220$ km/s, $v_{\mathrm{esc}}= 550\pm50$ km/s, the latter approximating the results from \cite{Piffl:2013mla}.   The total velocity of the dark matter relative to the Earth in galactic coordinates is
\begin{equation}
\vec{v}_{\chi}^{\,\mathrm{gal}}(t) = \vec{v}_{\mathrm{MB}} + \vec{v}_{\mathrm{LSR}} 
					+ \vec{v}_{\mathrm{pec}} + \vec{u}_\mathrm{E}(t)~.
\end{equation}
This expression takes into account the velocity of the local standard of rest $\vec{v}_{\mathrm{LSR}} = v_0\,\hat{y}^{\mathrm{gal}}$, the peculiar velocity of the Sun $\vec{v}_{\mathrm{pec}} = (11.1, 12.2, 7.3)$ km/s, and the Earth's velocity around the Sun $\vec{u}_\mathrm{E}(t)$, which varies over a sidereal year.  The combination of these velocities means that the dark matter wind comes from $\sim 47^\circ$ declination, inside the Cygnus constellation.  In determining the Earth's velocity around the Sun, we follow the procedure of Ref.~\cite{McCabe:2013kea}.  Combining these relative motions gives the net velocity of dark matter relative to the Earth in the galactic coordinate system.  However, one must also know the position of the scatter site and the lab in the same coordinate system as the velocity, which requires transforming between galacto-centric coordinates and Earth-centric (also called equatorial) coordinates.  The two frames are related by a series of rotations, $R_i(\theta)$, through angle $\theta$ around axis $i$, 
\begin{equation}
 \vec{v}_{\chi}^{\,\mathrm{gal}} = R_y(\theta_{\odot}). R_x(\eta).R_z(\alpha_{\mathrm{GC}}).R_y(\delta_{\mathrm{GC}}).\vec{v}_\chi^{\,\mathrm{equ}}.
\end{equation}
 Here, the right ascension of the galactic centre (GC) is $\alpha_{\mathrm{GC}}=266^\circ$, the declination is \mbox{$\delta_{\mathrm{GC}}=-29.0^\circ$}, $\eta=58.6^\circ$, and $\theta_{\odot}$ is determined by the height of the Sun above the galactic midplane, $\sin\theta_{\odot}=z_\odot/d_{\mathrm{GC}}$.  Of course, the magnitude of the velocity is unchanged under rotation and we denote the speed as $v_\chi$.

Now that we have determined the incoming velocity in the lab frame, we turn to the kinematics of the initial scatter.  Although we assume the scattering cross section is isotropic in the center of mass frame, the outgoing excited state will be forward, for the masses of dark matter we consider.  The kinematics in the center of mass frame are straightforward and the outgoing speed in this frame is given by
\begin{equation}
v_{\mathrm{out}}^{\mathrm{cm}}=\left[\frac{\mu_2}{m_2^2}\left(\mu_1 v_\chi^2-2\delta\right)\right]^{1/2}~,
\end{equation}
with $\mu_{(1,2)}=m_{(1,2)} m_T/(m_{(1,2)}+m_T)$ the reduced mass of the target and the incoming and outgoing dark matter.
The corresponding outgoing speed in the lab frame satisfies
\begin{equation}
v_{\mathrm{out}}^{\mathrm{lab}} = \left[(v_{\mathrm{out}}^{\mathrm{cm}})^2 + \left(\frac{\mu_1 v_\chi}{m_T}\right)^2 + \frac{2\mu_1 v_\chi v_{\mathrm{out}}^{\mathrm{cm}}}{m_T} \cos\theta^{\mathrm{cm}} \right]^{1/2}~.
\end{equation}
The scattering angle in the lab frame is related to the scattering angle in the center of mass frame through
\begin{equation}
v_{\mathrm{out}}^{\mathrm{lab}} \cos\theta^{\mathrm{lab}} = v_{\mathrm{out}}^{\mathrm{cm}}\cos\theta^{\mathrm{cm}} + \frac{\mu_1 v_\chi}{m_T}~.
\end{equation}
This angle is limited kinematically, with a maximum value $\theta_{\mathrm{max}}^{\mathrm{lab}}$ which satisfies
\begin{equation}
\cos^2\theta_{\mathrm{max}}^{\mathrm{lab}} = 1-\left(\frac{m_T v_{\mathrm{out}}^{\mathrm{cm}}}{\mu_1 v_\chi}\right)^2 = \left(1 + \frac{m_T}{m_{2}}\right)
		  \left(1 - \frac{m_T}{m_{1}}
			    + \frac{2\,m_T\,\delta}{(m_{1}\,v_{\chi})^2}\right)~.
\label{eq:costhetamaxiDM}
\end{equation}
The minimum velocity (in the lab frame) required to upscatter is given by $v_{\mathrm{min}} = \sqrt{2\,\delta/\mu_1}$; at large values of $\delta$, $v_{\rm{min}}$ will exceed the largest allowed value of $v_\chi$, driving the signal rate to zero.

The scattering angle necessary to reach the detector, $\theta^{\mathrm{lab}}$, must lie within the cone subtended by opening angle $\theta_{\mathrm{max}}^{\mathrm{lab}}$, and the fraction of the cone that the detector covers is given by \mbox{$[R_D/(|\vec{r}_s-\vec{r}_D| \theta_{\mathrm{max}}^{\mathrm{lab}})]^2$}, where $R_D$ is the radius of the detector; for example, for Borexino $R_D=5.5$ m (see \tabref{tab:labs}). There are two possible center of mass frame scattering angles that will result in the excited dark matter arriving at the detector, leading (in the lab frame) to two different outgoing speeds for the excited dark matter,
\begin{equation}
 v_{\mathrm{out},\pm}^{\mathrm{lab}} = \frac{m_{1}\,v_{\chi}}{m_{2} + m_T} \cos\theta^{\mathrm{lab}}
	  \left[1 \pm \sqrt{1 - \frac{\cos^2{\theta^{\mathrm{lab}}_{\mathrm{max}}}}{\cos^2{\theta^{\mathrm{lab}}}}}\right].
\end{equation}
These two solutions take different lengths of time to get from the scatter site to the detector.  The probability that an excited state moving at speed $v_\mathrm{out}^{\mathrm{lab}}$ will travel a distance $L = \left|\vec{r}_s - \vec{r}_D \right|$ and decay inside the detector is
 \begin{equation}
 P(v_\mathrm{out}^{\mathrm{lab}}, L, \tau) = 2\,\sinh\Big(\frac{R_D}{2\,v_\mathrm{out}^{\mathrm{lab}}\,\tau}\Big)\exp\left(-\frac{L}{v_\mathrm{out}^{\mathrm{lab}}\tau}\right)~.
 \end{equation}
Similarly, for the two possible center of mass scattering angles there are two different values of momentum exchanged with the nucleus, resulting in two different form factor suppressions.  To account for the substructure of the nucleus, we use the Helm form factor\cite{Lewin:1995rx}
\begin{equation}
 F(q) = \frac{3}{q\,r_n}\,J_1(q\,r_n)\,\exp\left(-\frac{q^2 s^2}{2}\right)~,
\end{equation}
with $s\approx 0.9$ fm and $r_n \approx 1.14\,(A/0.93149)^{1/3}$ fm, and where $q = \sqrt{2\,m_T\,E_{R}}$ is the momentum transfer in the collision. 

In the center of mass frame the cross section is isotropic and almost independent of the incoming velocity,
 \begin{equation}
 	\frac{d\sigma}{d\cos\theta^{\mathrm{cm}}} = \sigma_0 \,
	    \sqrt{1 - \frac{2\,\delta}{\mu_1\,v_{\chi}^2}}
	    \left(\frac{m_{1}}{m_{1}+m_T}\right)^2\,A^4~,
 \end{equation}
where $A$ the mass number of the target nucleus. To determine the fraction of scatters that end up in the detector, we must integrate over all scattering angles in the lab frame.  The transformation to the lab frame introduces a Jacobian  
 \begin{equation}
   J_\pm(v_\chi) = \frac{d\cos\theta^{\mathrm{cm}}}{d\cos\theta^{\mathrm{lab}}} \nonumber \\
      = 2\,z\,\cos\theta^{\mathrm{lab}} \pm 
	      \frac{1 - z^2 + 2\,z^2\,\cos^2\theta^{\mathrm{lab}}}
		    {\sqrt{1 - z^2 + z^2\,\cos^2\theta^{\mathrm{lab}}}}~,
\end{equation}
with $z= \mu_1 v_\chi/(m_T\, v_{\mathrm{out}}^{\mathrm{cm}})$.

Putting all these effects together we arrive at the final result for the expected rate inside the detector,
\begin{align} \label{eq:totalrate}
 \Gamma = \sum_{\pm}\int d^3r_s\,d^3v_{\mathrm{MB}}\,
 		\Big\{n_T(r_s)\,\frac{\rho_\chi}{m_{1}} 
		&\left[\frac{R_D}{\left|\vec{r}_s - \vec{r}_D\right|\,\theta_{\mathrm{max}}^{\mathrm{lab}}}\right]^2\,
		P(v_{\mathrm{out},\pm}^{\mathrm{lab}}, L, \tau)\nonumber \\
        &\times f_{\mathrm{gal}}(v_{\mathrm{MB}})\, \left|F(q_\pm)\right|^2\,
        \frac{d\sigma \,v_{\chi}}{d\cos\theta^{\mathrm{cm}}}\left|J_\pm(v_\chi)\right|^2\,
        		 \Big\}~.
\end{align}
In the results presented below we have evaluated this integral numerically.

\subsection{Results for modulating rates}
\label{sec:resultsmodulatingrates}

We now present the results for the signal rate as a function of the time of day in a few representative experiments. We consider three different lab locations: Gran Sasso in Italy, Jiangmen in China, and SUPL  in Australia. Gran Sasso is home to the Borexino detector~\cite{Bellini:2013lnn} and is also one of the possible locations for CYGNUS~\cite{SpoonerIDM:2018,CYGNO:2019aqp}, a large gas TPC for directional dark matter detection. Jiangmen will host the JUNO neutrino detector~\cite{Djurcic:2015vqa}, which will be be notably larger than Borexino. We also consider a hypothetical Borexino-like detector at SUPL\@. The lab locations and detector sizes are shown in~\tabref{tab:labs}.
\begin{table}[tb]
\begin{center}
\begin{tabular}{cccccc}
 \hline \hline
 Lab & Detector  & latitude  & longitude & depth [km] & Radius [m]   \\
  \hline 
   Gran Sasso & Borexino  & $42.7^\circ$ N & $13.6^\circ$ E & $1.4$ & 5.5 \\
   Gran Sasso & CYGNUS & $42.7^\circ$ N & $13.6^\circ$ E & $1.4$ & 1-10 \\
 Jiangmen  &  JUNO  & $22.1^\circ$ N & $112.5^\circ$ E & $0.48$  & 17 \\
  SUPL & hypothetical & $37.1^\circ$ S & $143^\circ$ E & $1.0$  & 5.5 \\
  \hline \hline
\end{tabular}
\end{center}
\caption{The position, depth, and size of various underground detectors, both existing and proposed, which are analyzed in this work. SUPL is the Stawell Underground Physics Laboratory in Australia, and we place a hypothetical Borexino-like detector there.
}
\label{tab:labs}
\end{table}

\begin{figure}[!th]
\begin{center}
 \includegraphics[scale=.53]{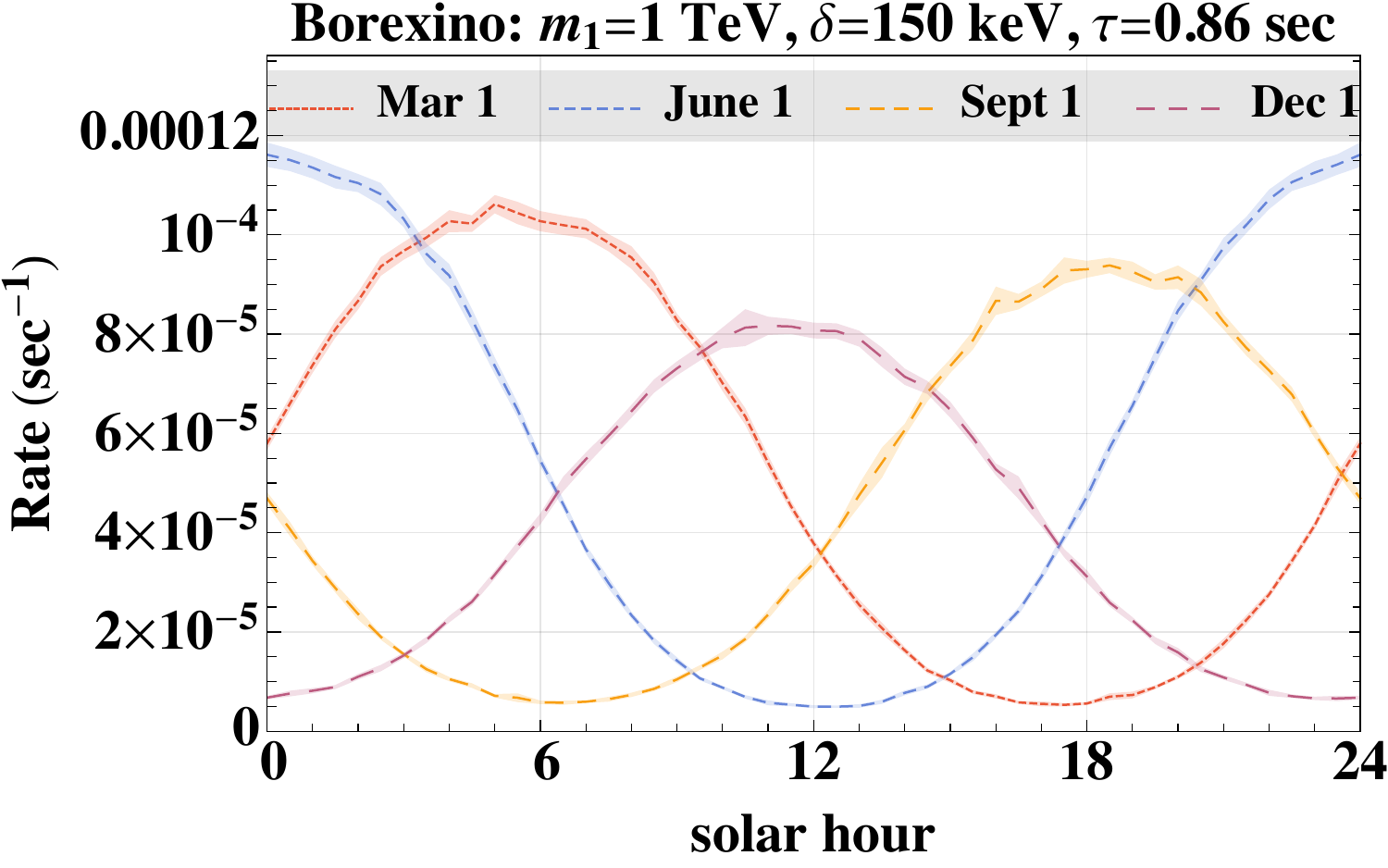}
 \includegraphics[scale=.53]{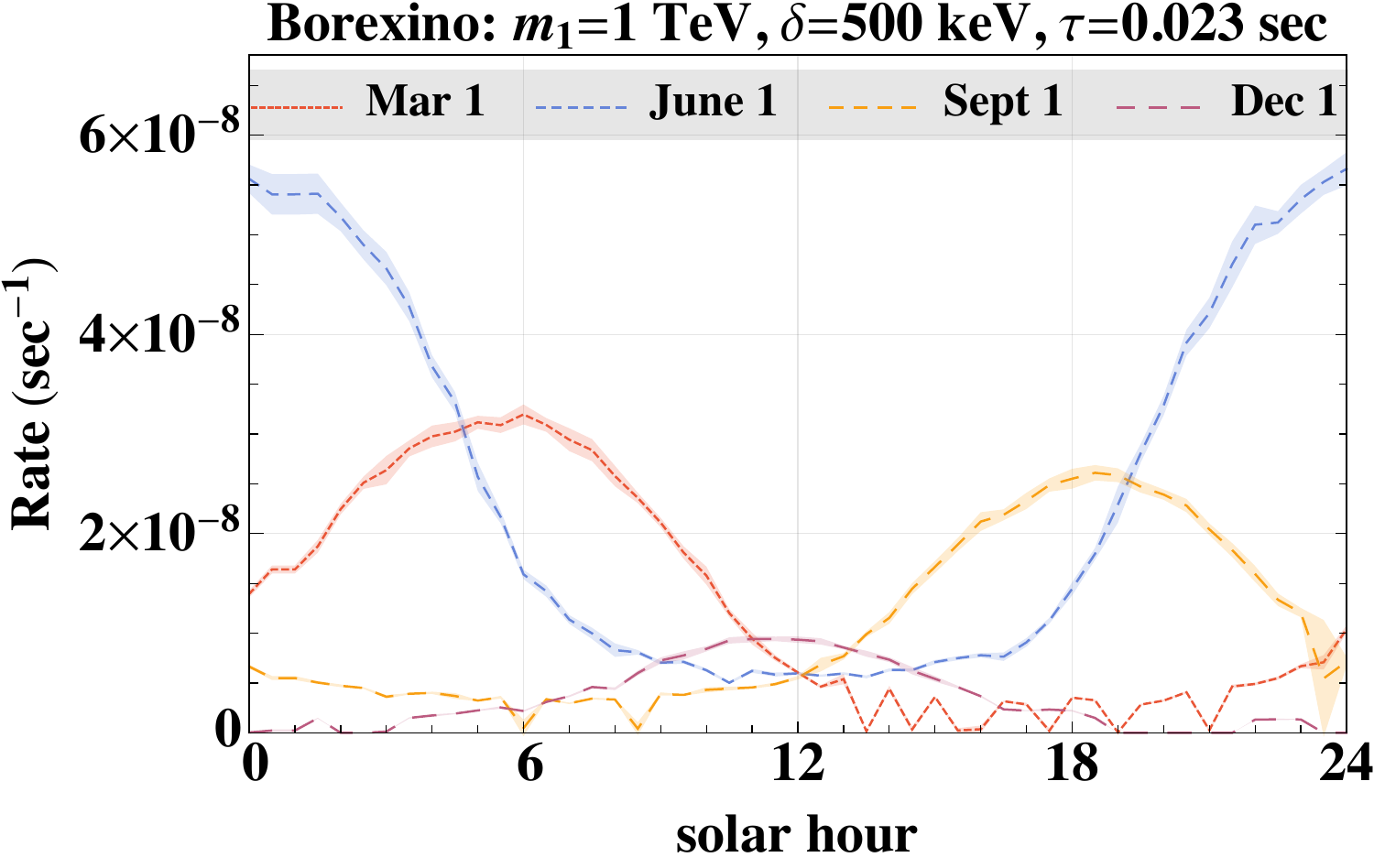} \\ \vspace{0.1cm}
 \includegraphics[scale=.53]{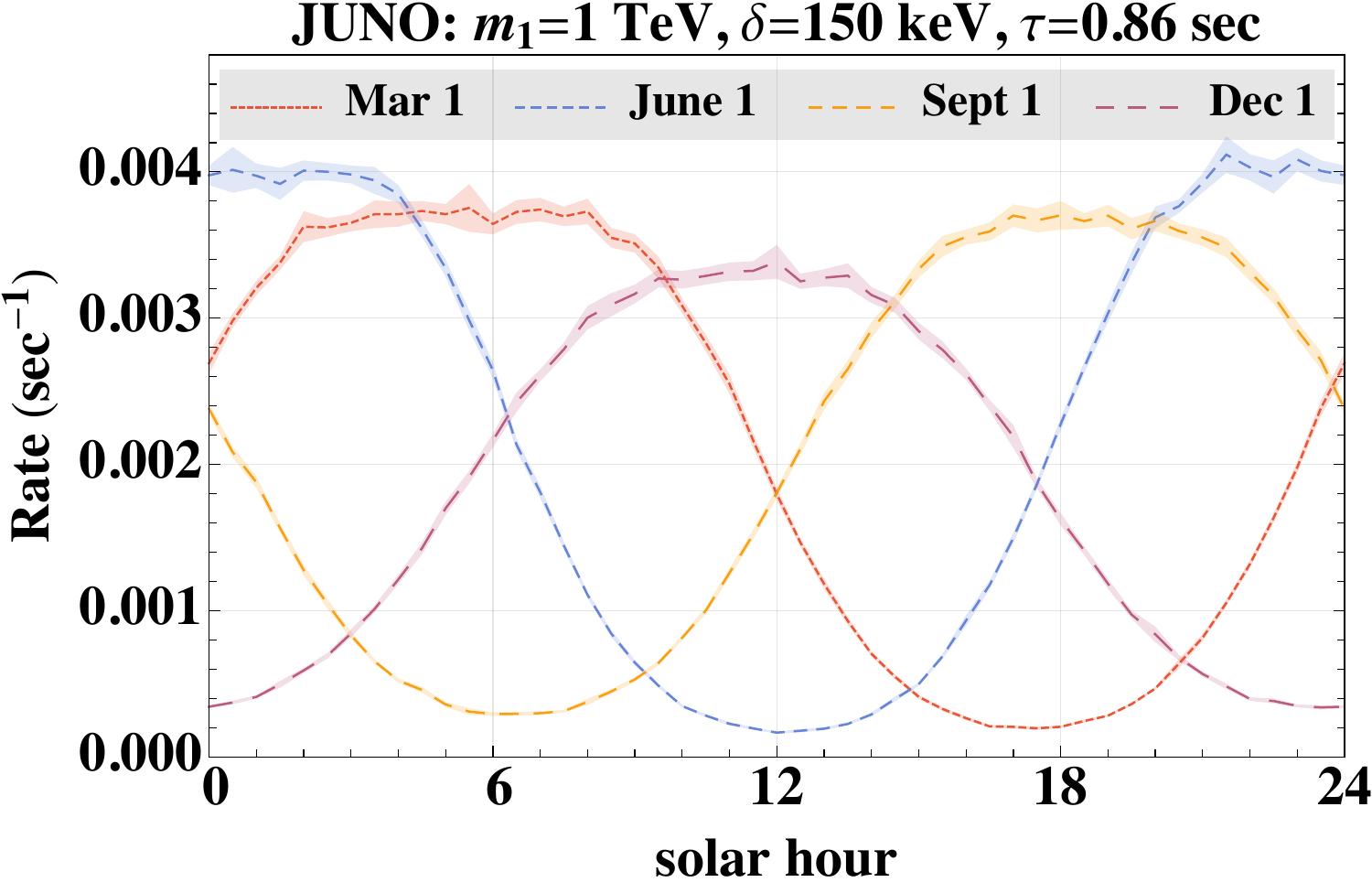} 
 \includegraphics[scale=.54]{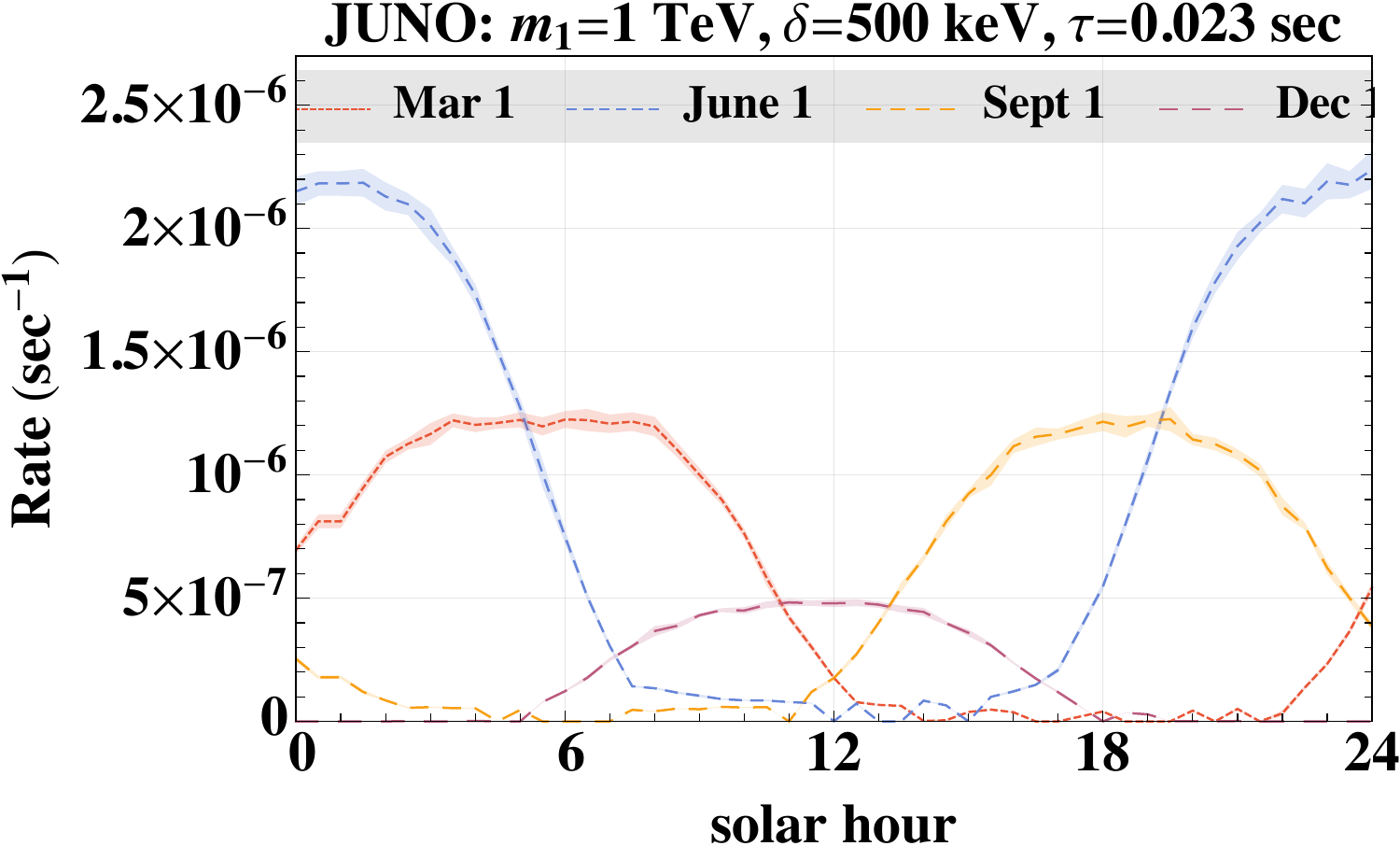} \\ \vspace{0.1cm}
 \includegraphics[scale=.53]{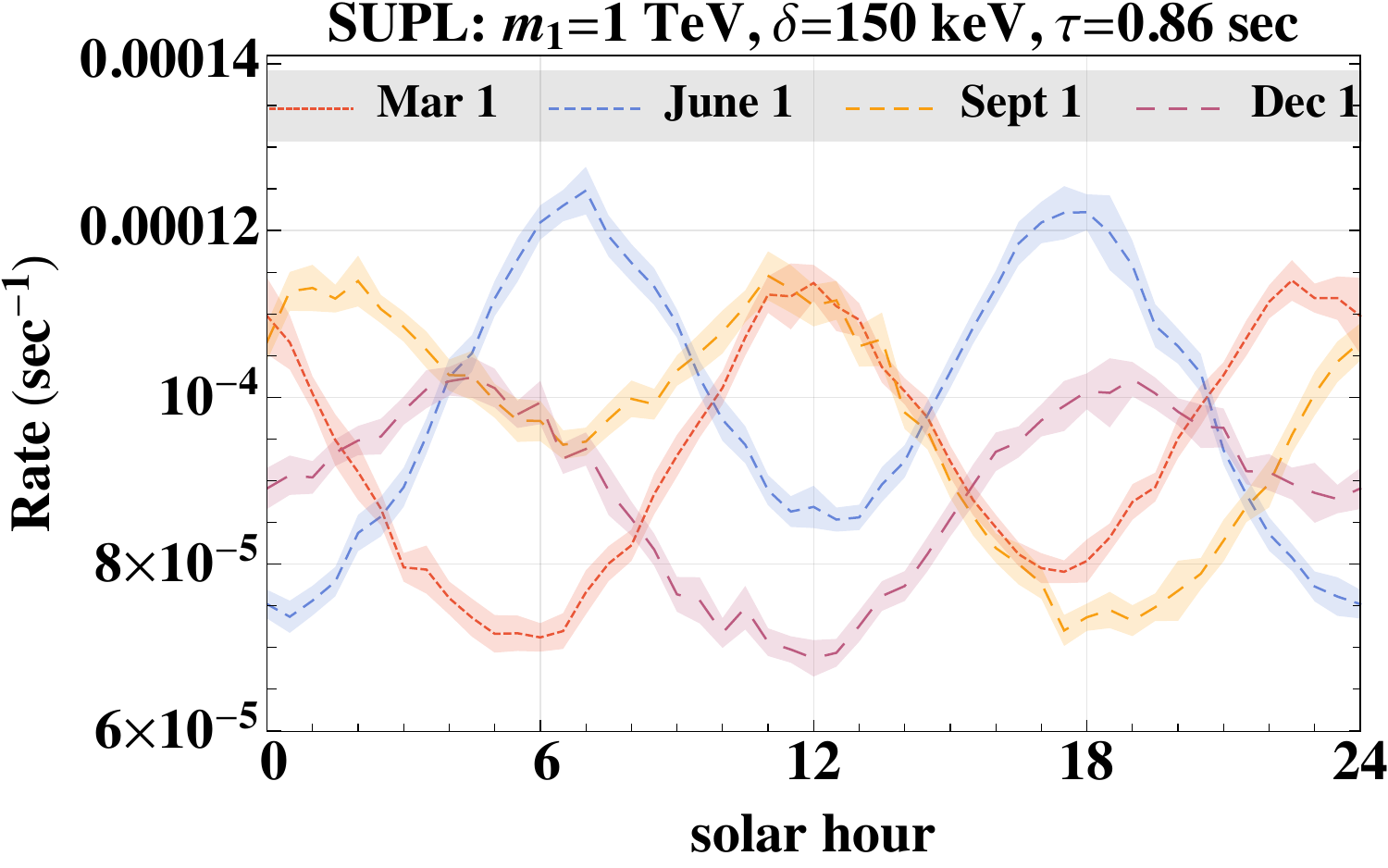}
 \includegraphics[scale=.53]{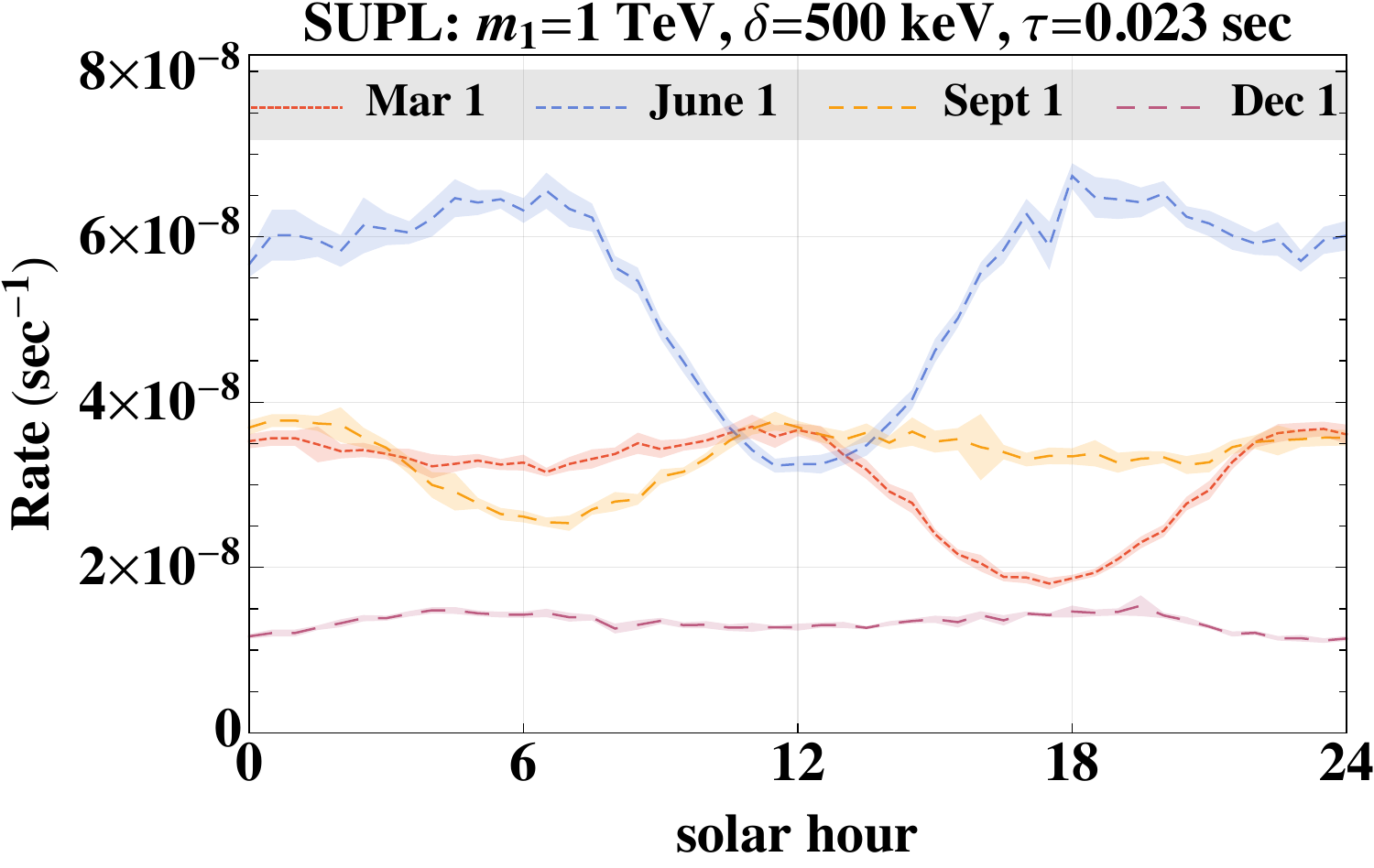}
\end{center}
 \caption{The expected signal rates for the Borexino detector (top row), JUNO detector (middle row), and a hypothetical detector at SUPL (bottom row).  The curves show the rate over one solar day: the first day of March (red), June (blue), September (yellow), and December (purple). The rate was determined using \eref{eq:totalrate} with $m_\chi = 1$ TeV and lifetime given by the decay length of \eref{eq:decaylength}; the mass splitting in the left (right) panels is $\delta = 150$ ($500$) keV\@. The results are shown in a common time-zone for all labs for ease of comparison.}
 \label{fig:modulation}
\end{figure}

The rates calculated  using \eref{eq:totalrate} are shown in \figref{fig:modulation} for TeV dark matter with mass splittings of 150 and 500~keV for the three liquid scintillation detectors we consider. 
For the lifetime of the excited state, we use 
\eref{eq:decaylength} for a higgsino (this assumption will be relaxed below).

The curves show the rate (in events/s) over a period of one day: 
the first day of March (red), June (blue), September (yellow), 
and December (purple). The first feature to notice is that the rate modulates strongly, as expected. Secondly, the time of day of the peak rate changes throughout the year. This is because the rate modulates with a period of a sidereal day, which is approximately four minutes shorter than the solar day. It is interesting to note that our modulation effect would be washed out if one were searching for a (solar) day-night asymmetry with an exposure of several years.  These distinct modulation patterns can be used to discriminate signal from backgrounds, as we will show in the next section.
The modulation patterns in \figref{fig:modulation} have some interesting characteristics: 
\begin{itemize}
\item For high $\delta$ the rate modulates significantly both with a period of a sidereal day and annually. The later is due to the usual enhanced modulation of inelastic dark matter. 
\item In comparing the rate at Borexino and JUNO one can see that the peaks of an enhanced rate are wider in JUNO\@. This is because it is located further south, where Cygnus spends more of the day below the horizon.
\item In all of the plots, we see the peak rate shifts by approximately 6 solar hours from season to season due to the difference between the solar and the sidereal day durations.  
\item In comparing the left and right in the top two panels of \figref{fig:modulation}, we notice the  the transition from high to low rate occurs more sharply for high $\delta$ as compared to low. This is because large splitting requires faster incoming dark matter. As shown in the top of \figref{fig:fluxCygnus}, the dark matter particles are coming from a more focused region in the sky. The smaller spot takes less time to set below the horizon leading to a faster transition.
\item In the Southern Hemisphere, the Cygnus constellation does not rise far above the horizon, as a result, the rate at SUPL does not drop as close to zero as compared to the Northern Hemisphere detectors.
\item Notice that there are some interesting doubly-peaked modulation patterns in the Southern Hemisphere. These can be understood by noticing that the Earth's crust is richer in lead, than is the mantle. Inspecting \figref{fig:apparentdepths} ones sees that the crust contribution to the apparent depth peaks at two distinct times during the sidereal day: once when Cygnus sets and once when it rises.
\item While the event rate per day can be very small, a modulating signal can be seen above background by stacking across multiple days. This requires a long exposure of the experiment and/or a larger signal cross section, as we will show in~\secref{sec:sensitivity}.
\end{itemize}

\begin{figure}[t]
\includegraphics[scale=.38]{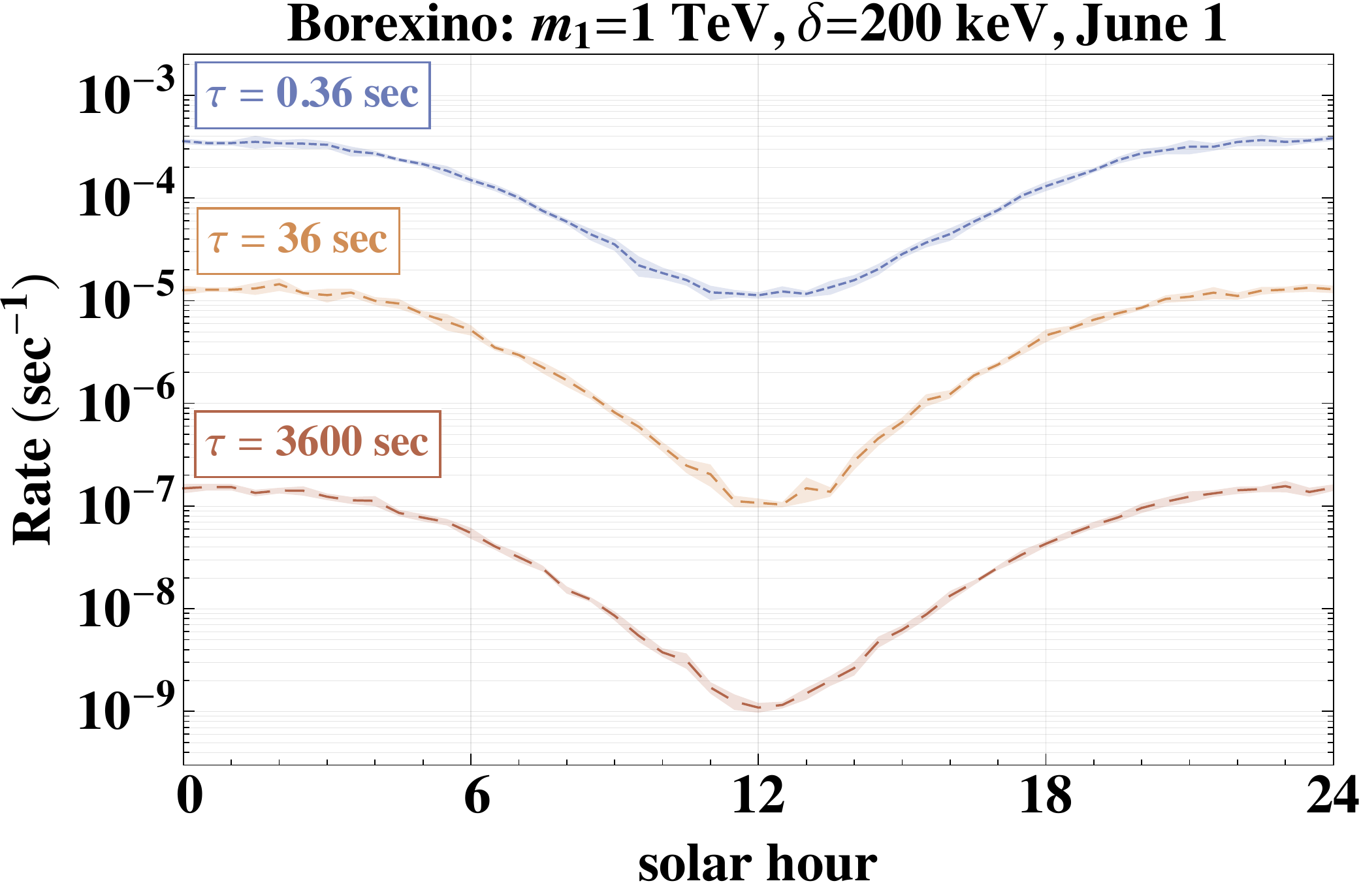}
\includegraphics[scale=.38]{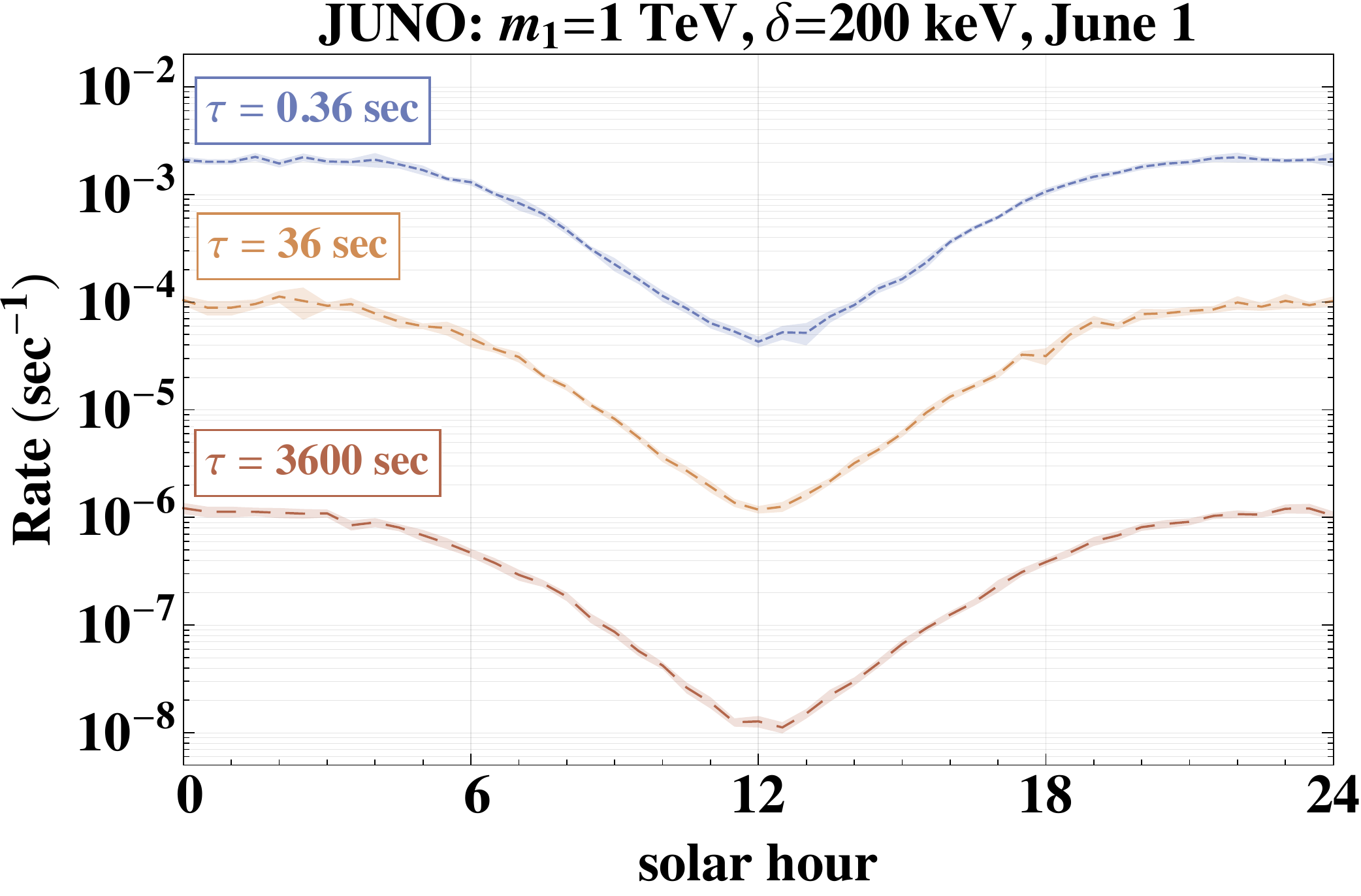} \\
\includegraphics[scale=.38]{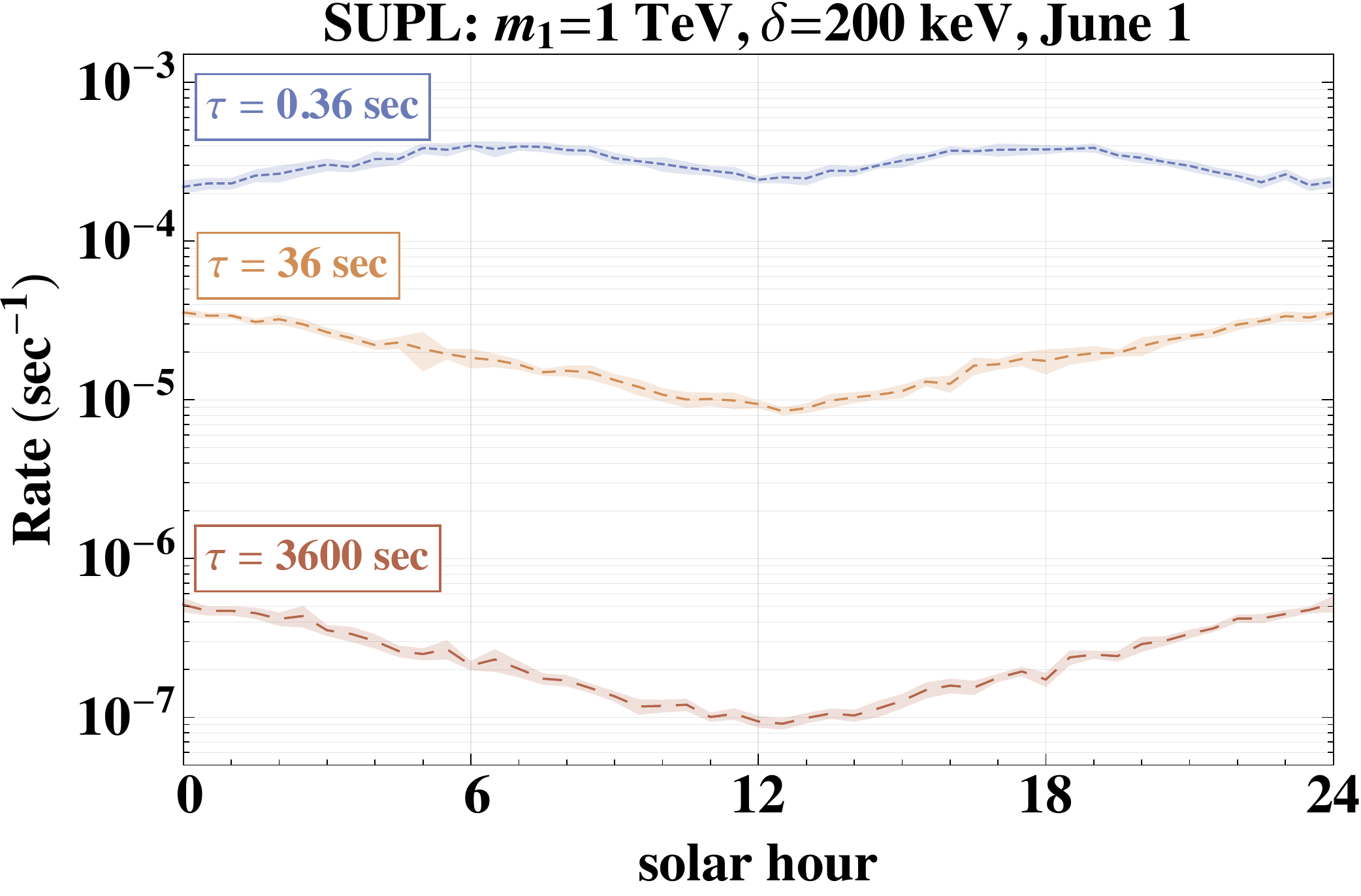}
\includegraphics[scale=.38]{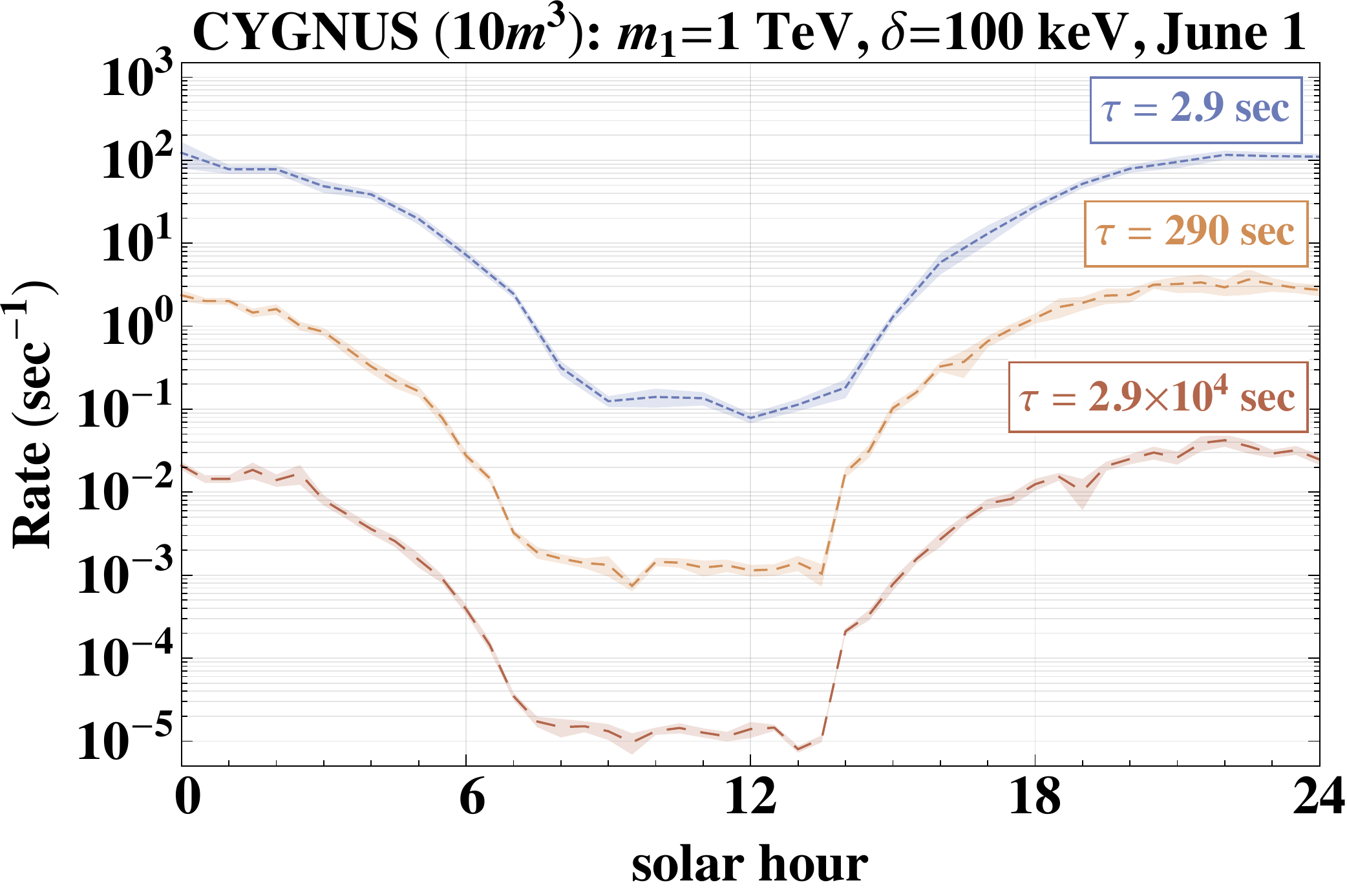}
 \caption{The total signal rate at $\delta=200$ keV over one day for Borexino (top left), JUNO (top right), and SUPL (bottom left), and the rate for $\delta=100$~keV for a $10$m$^3$ CYGNUS-like detector (bottom right); the lifetime is taken to be $\tau = \ell_{\chi_2^{}}/c$ (blue), $10^2 \ell_{\chi_2^{}}/c$ (yellow), and $10^4 \ell_{\chi_2^{}}/c$ (red), where $\ell_{\chi_2^{}}$ is given by \eref{eq:decaylength}.}\label{fig:modulation-lifetimes}
\end{figure}

The fact that the signal modulates is expected to be quite robust to changes in the lifetime of the excited state $\chi_2$, although the quantitative details will vary with lifetime. This is because, as the lifetime grows, the effective volume for scattering that can reach the detector grows, while the angular acceptance of the added volume is correspondingly smaller. However, as the lifetime of the excited state grows, different parts of the Earth contribute. In addition, if the lifetime exceeds the crossing time of the Earth, the signal rate begins to decrease linearly as $v\tau/R_\oplus$. 
To demonstrate this, in \figref{fig:modulation-lifetimes} we plot the modulation rate in several detectors for varying $\chi_2$ lifetimes. The linear drop in event rate is seen clearly. In addition, the  differences in shape between lifetimes of $0.36$ seconds and longer lifetimes is because the former is sensitive to the Earth composition closer to the detector whereas longer lifetimes probe the whole Earth. For example, the double peak structure in the Southern Hemisphere disappears when the whole Earth is probed. 

\section{Current and Future Sensitivities}
\label{sec:sensitivity}

We are now ready to estimate the sensitivity of modulating luminous signals to inelastic dark matter. We will focus on two regions of parameter space which can be probed by two different types of detectors. In \secref{sec:borexino}, we consider higher mass splitting, $\delta\gtap 200$~keV, which can be probed by detectors with a high threshold, such as the large liquid scintillator neutrino detectors Borexino and JUNO\@.   In \secref{sec:gas}, we will consider low mass splittings which can be probed by large gaseous detectors designed for dark matter directional detection, such as the proposed CYGNUS detector.  

\subsection{Sensitivity of Borexino and JUNO}
\label{sec:borexino}

\begin{figure}[t]
\begin{center}
\includegraphics[width=0.65\textwidth]{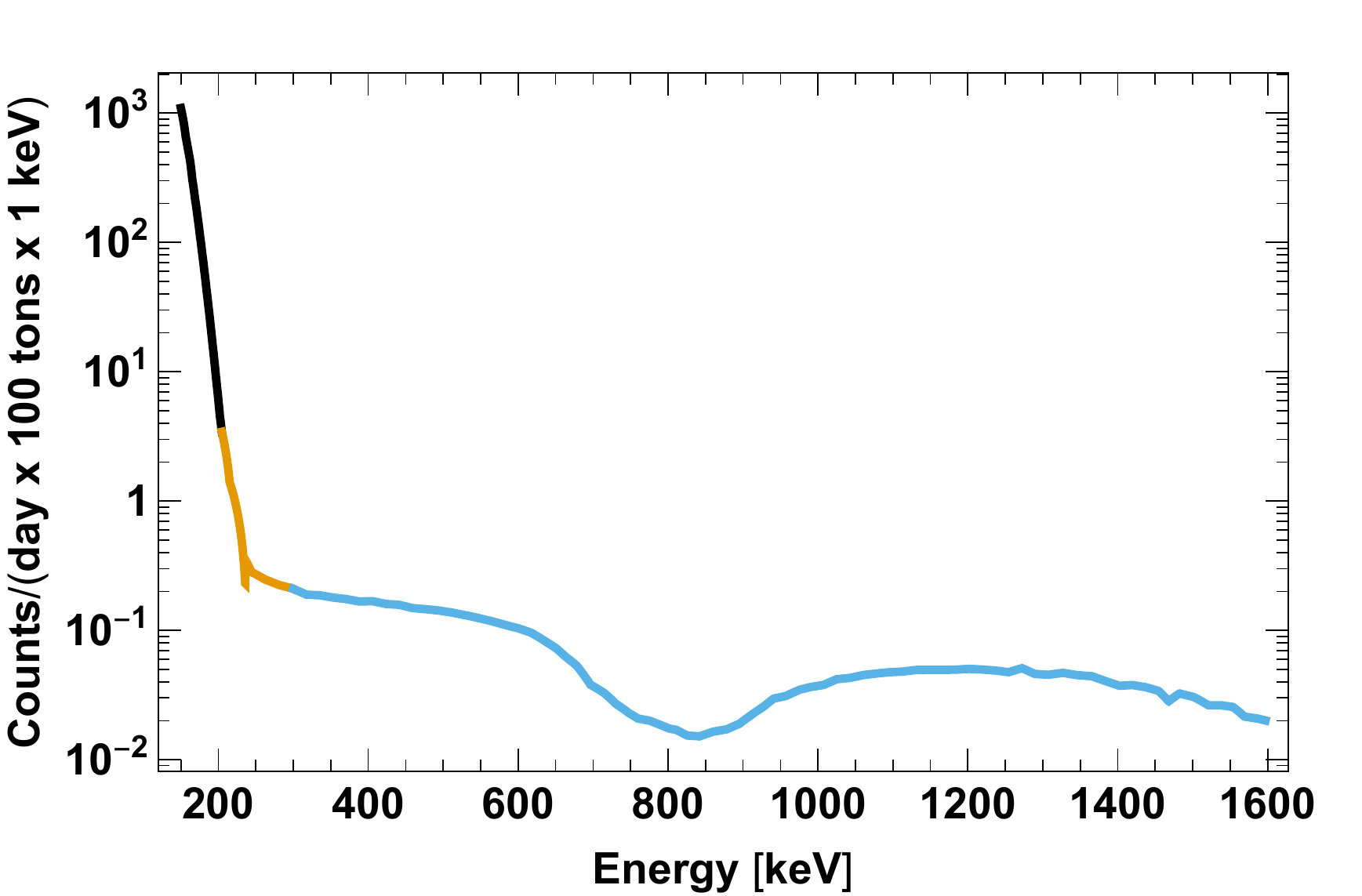}
 \caption{The background rate at Borexino.  The black curve is extracted from \cite{Agostini:2015oze} while the orange and blue curves come from \cite{Bellini:2013lnn}.}\label{fig:backgrounds}
 \end{center}
\end{figure}

In this section, we consider the Borexino detector and events 
observed during  
its running of $1291.5$ days~\cite{Bellini:2013lnn}. The observed rate in the vicinity of 250-600~keV is
$0.1-0.5$~events/(day$\times$keV$\times$100~tonnes). This rate does not include the radioactive background from $^{210}$Po $\alpha$-decay, since it can be effectively subtracted using a fit of shower shape variables (see e.g.\ Figure~53 of~\cite{Bellini:2013lnn}). The rate is dominated by the $^7$Be neutrino signal, which amusingly serves as a background for our analysis.  There are also several radioactive backgrounds which play a significant role. Below $250$~keV the $^{14}$C background dominates, producing a much higher background rate. 
The background rate that we use to place our bounds is shown in \figref{fig:backgrounds}, and was extracted from~\cite{Agostini:2015oze} and~\cite{Bellini:2013lnn}. 
In our estimate we will \emph{not} assume any fundamental understanding of these backgrounds but will instead make use of the daily modulation of the signal. 

The luminous dark matter signal is a spectral line at an energy $E_\gamma \simeq \delta$ on top of this background. This line is smeared by the energy resolution which we take to be $\sim 10\%$~\cite{Bellini:2013lnn}. The background rate within the line width is thus of order $5$~events per day in Borexino for photon energies above $250$~keV\@. 
If it were not for the daily modulation of the signal, a conservative bound would require that the luminous photon signal rate be below the observed background rate. This would not make use of the large exposure of Borexino. However, the limit obtained by a modulation analysis is much stronger and does take advantage of the order thousand tonnes$\times$year exposure of Borexino.

Performing a full-fledged modulation analysis is beyond the scope of our work.
We instead follow a simpler (though cruder) approach -- dividing the sidereal day into two bins: signal-``on'' and signal-``off''. For simplicity we take each bin to be one-half of a sidereal day. The ``on'' bin consists of the half-day in which the integrated signal rate is maximal, while the ``off'' bin is the other half-day in which the signal rate is lowest. In most of the parameter space of interest, the signal rate during the on bin will be much larger than that in the off bin, $\Gamma_s^\mathrm{on}\gg \Gamma_s^\mathrm{off}$ and the off rate can be neglected\footnote{In the off bin the signal is not quite off, but the validity of this approximation justifies its name.}. In the Northern Hemisphere the on bin consists of twelve consecutive sidereal hours, but this does not have to be the case in general as can be seen in \figref{fig:modulation}.

Employing this on-off approximation, one can ``measure'' the backgrounds using the signal-off bin and use this to help search for the signal in the signal-on bin. The uncertainty in the prediction of the background in the on bin will thus be set by the statistical uncertainty of the measurement in the off bin.  
Suppose that during Borexino's full run it has accumulated $N_\mathrm{off}$ events during signal-off times in a window of size $\pm 0.1\delta$ around a hypothesized $\delta$. 
The reach of Borexino can be estimated by requiring that the signal rate in the signal-on bin, $\Gamma_\mathrm{signal}$, does not exceed the expected statistical fluctuations in the background rate in the signal-off bin,
\begin{equation}
\Gamma_\mathrm{signal} 
\ltap 
\frac{2\times 1.64}{\sqrt{N_\mathrm{off}}}\Gamma_\mathrm{off} 
\end{equation}
where $N_\mathrm{off} = \Gamma_\mathrm{off}\, t_\mathrm{off}$  and $ t_\mathrm{off} =1/2 \times 1291.5$ days is the total signal-off time accumulated in the exposure, and the factor of $1.64$ arises because we consider a $90\%$ confidence interval.
This procedure yields an expected limit which strengthens as the square root of the exposure.

\begin{figure}[t]
\begin{center}
\includegraphics[scale=.6]{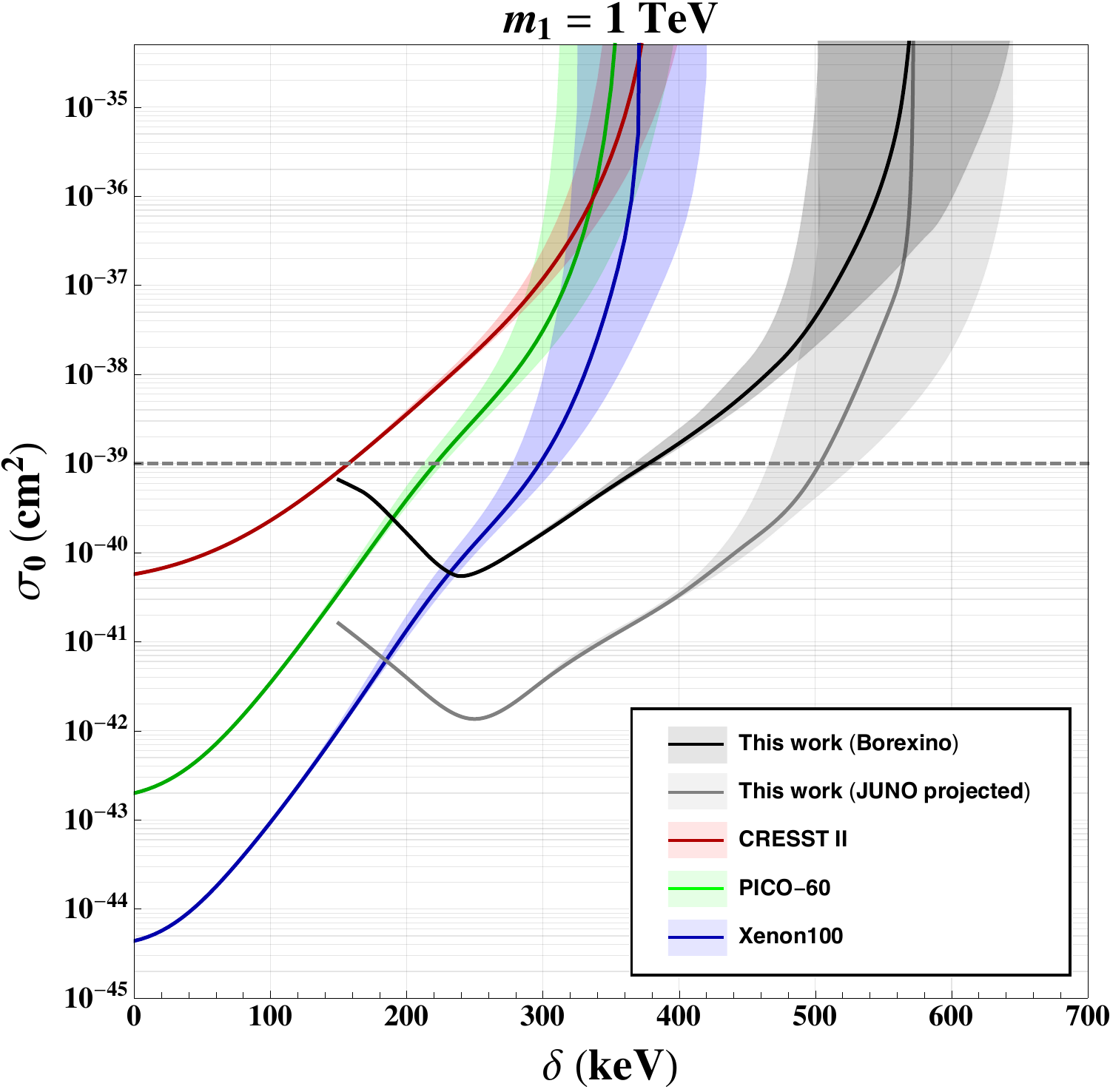}
 \caption{Projected sensitivity to the inelastic dark matter cross section in neutrino detectors. The red, green, and blue curves represent (respectively) existing direct detection bounds from CRESST~II, \mbox{PICO-60}, and XENON100 (see \cite{Bramante:2016rdh}). The shaded regions of each curve represent the uncertainty in $v_{\mathrm{esc}} = 550 \pm 50$~km/s. The black (gray) curves represent projected Borexino sensitivity using the existing $1291.5$ days of running (projected JUNO sensitivity assuming Borexino-like background and run time). The dashed horizontal gray line is the scattering cross section for a narrowly-split higgsino.} \label{fig:sigma}
 \end{center}
\end{figure}

In \figref{fig:sigma} we show the sensitivity of Borexino for a $1$~TeV higgsino with decay length determined by \eref{eq:decaylength}.  
This figure is one of 
the main results of this paper. The figure demonstrates that Borexino (and, in the future, JUNO) has the potential to substantially extend the bound on narrowly-split higgsinos, $\delta \gtap 380$~keV ($\delta \gtap 500$~keV), using an alternative search strategy to direct detection experiments. The weakening of the bound for $\delta \ltap 250$~keV is due to the growth of the background at low energies, specifically $^{14}$C\@.  We expect that an actual analysis, using the time-stamps in the real data, will be able to improve upon our analysis at high $\delta$ and potentially look underneath this background which will not have a sidereal modulation.

The current bounds, also shown in \figref{fig:sigma}, from the non-observation of nuclear recoil from  inelastic scattering are derived from \cite{Bramante:2016rdh} and shown using the latest results from CRESST-II~\cite{Angloher:2015ewa}, PICO-60~\cite{Amole:2015pla}, and XENON100~\cite{Aprile:2017aas}.  
CRESST has accumulated $52$~kg$\times$days of data with nuclear recoil energies between $20$-$120$~keV$_\mathrm{nr}$, they observed $4$ events.  PICO's data corresponds to $\sim 1300$~kg$\times$days of exposure with a sensitivity to recoils between $10-10^3$~keV, the lower threshold varied over their data taking from $7-20$~keV but we take the lower limit fixed at $10$~keV, they did not observe any events.  The conventional analysis of XENON100~\cite{Aprile:2016swn} is over an energy range of $6.6$-$43.3$~keV$_\mathrm{nr}$. However, the collaboration have also carried out an analysis up to $240$~keV$_\mathrm{nr}$, corresponding to $3-180$~PE in S1, and have presented their $\sim 7600$~kg$\times$days of data over an enlarged energy range, up to $1000$~PE in S1 (see Figure~9 of Ref.~\cite{Aprile:2017aas}).  They have not carried out a complete analysis over this full range, so the efficiencies for nuclear recoils are not known; we assume they are similar to those below $180$~PE\@.  Since they have not seen any events up to $1000$~PE ($\sim 500$~keV$_{\mathrm{nr}}$), we place a bound.   In all cases, after accounting for the mass fraction of the experiment made of the heavy target element, we must apply a rescaling to account for efficiency effects.  We determine this rescaling by matching with the known constraint at $\delta=0$~keV\@.

The sensitivity to the photon signal in Borexino (and presumably JUNO, 
in the future) is limited by radioactive backgrounds and solar neutrinos 
scattering off the elements in the liquid scintillator.  The neutrino 
scattering was, after all, the motivation for the design of these 
experiments.  Substantial improvement could be achieved if these 
``backgrounds'' could be reduced.  A similar sized detector but with a 
significantly reduced mass inside the detector, achieved by swapping
the liquid scintillator with a gaseous scintillator, would yield significant
improvements in the sensitivity to luminous dark matter.
In the regime $\delta \gtap 150$~keV, a Borexino-sized detector would 
see nearly two orders of magnitude improvement in the sensitivity 
to the luminous dark matter signal.  Actually modifying the Borexino
detector to use a gaseous scintillator would be very interesting, but 
undoubtedly is accompanied by numerous experimental challenges. 
However, as we see in the next section, directional dark matter detectors
employ large volumes of gaseous scintillator that are ideal to reach 
sensitivity to a lower range of $\delta$.

\subsection{Sensitivity of gaseous scintillation detectors: CYGNUS}
\label{sec:gas}

The rate of the photon signal from decay of the excited state scales with the
\emph{volume} of the detector and not the \emph{mass} of the 
scintillation material.  Since the backgrounds to the luminous signal 
scale with mass, the simplest way to reduce, and potentially eliminate, 
the backgrounds would be to change the detection technology from a 
liquid scintillator to a gaseous scintillator material.  

Existing and proposed directional dark matter detectors
including DMTPC \cite{Deaconu:2017vam},
DRIFT \cite{Battat:2016xxe} and CYGNUS 
\cite{SpoonerIDM:2018,CYGNO:2019aqp}
employ a gaseous scintillation material.
The sensitivity of these detectors to 
ordinary elastic dark matter scattering is limited by
the mass of the scintillator, which is why they have 
primarily focused on gaseous elements sensitive to 
spin-dependent elastic scattering.  The luminous dark matter
signal, however, does not depend on scattering within the
detector volume and is thus independent of the 
particular gas used.  The luminous signal only requires good efficiency 
in converting the $\mathcal{O}(10-500)\mathrm{\,keV}$ photon into 
scintillation light.

One of the principal benefits of the gaseous scintillator
detectors is their low detection threshold for electromagnetic-equivalent energy deposition,
of order few to $10$~keV$_{ee}$, which is considerably lower than the threshold for
neutrino detectors.  In addition, some of the choices for the scintillation
gas do not involve carbon (instead, SF$_6$ or He), which should 
decrease the radioactive background from $^{14}$C\@.  Hence, gaseous 
scintillators can be sensitive to much lower energy photons coming from 
small mass differences, $\delta \gtap$~few to $10$~keV, between 
the excited state and the dark matter.

\begin{figure}[t]
\begin{center}
\includegraphics[scale=.6]{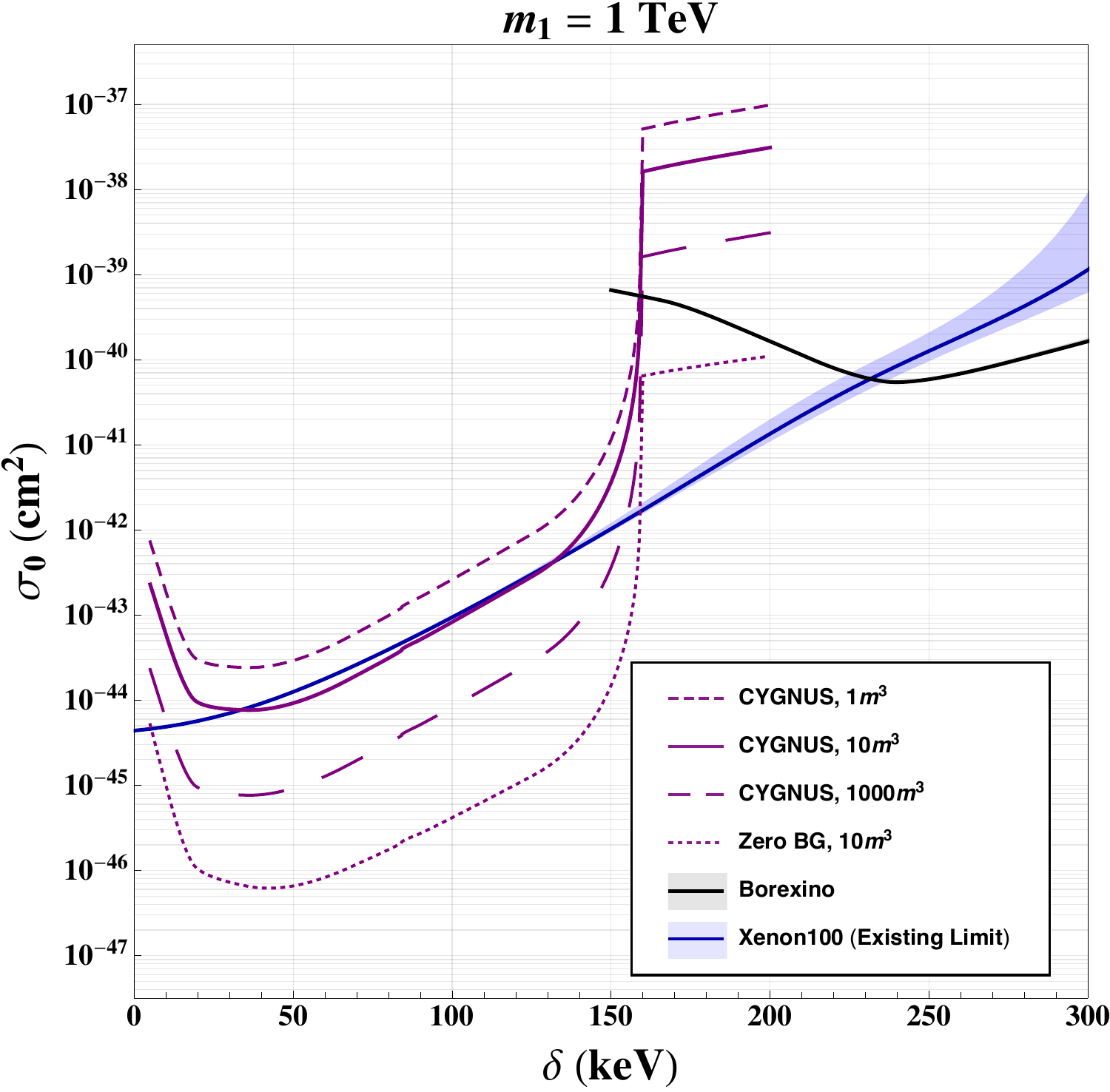}
 \caption{Projected sensitivity to the inelastic dark matter cross section in gaseous dark matter detectors such as CYGNUS\@. The blue curve represents existing direct detection bounds from XENON100, and the black solid line is the projected limit from Borexino, as in \figref{fig:sigma}. 
 The purple lines represent an experiment with \mbox{CYGNUS-like} background projections, with detector volume $1$~m$^3$ (short dashed), $10$~m$^3$ (thick), and $1000$~m$^3$ (long dashed). The dotted purple line represents the projected sensitivity of a hypothetical zero-background experiment with detector volume $10$~m$^3$. For the CYGNUS curves, we only show the sensitivities with $v_{\mathrm{esc}} = 550$~km/s.} \label{fig:sigmaCYGNUS}
 \end{center}
\end{figure}

Lower $\delta$ means lighter and more abundant
elements in the Earth can be used to upscatter.  While there are 
many elements lighter than lead that could be used to scatter,
in the following we illustrate the sensitivity of the CYGNUS 
detector using upscattering off iron and silicon (as well as lead).  
Iron is the fourth most abundant 
element by mass in the Earth and has a number density that is 
$10^{5}$-$10^{7}$ times larger than lead.  Silicon is the second most abundant 
and has a number density a further $10$ times larger than iron in the 
crust and the mantle (but $10$ times smaller in the core), see \tabref{tab:Density}.
Upscattering is kinematically possible off iron with 
$\delta \ltap 160$~keV and possible off silicon for $\delta \ltap 85$~keV, 
which means gaseous detectors can probe a complementary range of 
inelastic parameter space.  

In \figref{fig:sigmaCYGNUS}, we show the sensitivity
that the CYGNUS detector, assumed to be sensitive to photon energies $5\,\mathrm{keV}\le E_\gamma \le 200\,\mathrm{keV}$, would be capable of reaching with several
potential volumes ($1$, $10$, and $1000$~m$^3$) after one year of exposure.  Since these detectors have yet to be built, the background rate is not yet known.  Estimates for a 10~m$^3$ detector in the 1-10 keVee energy range have been presented in \cite{SpoonerIDM:2018}. We extrapolated these estimates above the 1-10 keVee energy window, assuming they are independent of energy, up to $200$ keV and also assumed the background rates scale with detector volume for the larger possible detectors.  We do not estimate the sensitivity for $\delta>200$ keV since this is typically the highest energy at which existing directional detector collaborations have calibrated their detector response.

The dramatic increase in sensitivity 
below $160$~keV is a result of inelastic dark matter being able to 
scatter off iron.  In practice, there would be improved sensitivity
even between ${150 \text{ keV} \ltap \delta \ltap 200 \text{ keV}}$ due to elements 
in the Earth that are more abundant than lead but also heavier than 
iron (such as barium).  For $\delta \ltap 85$ keV it is possible for dark matter to scatter off silicon, and for lower $\delta$ this will dominate. Once $\delta \ltap 40$~keV,
the characteristic decay 
length, \eref{eq:decaylength}, exceeds the diameter of the Earth and the sensitivity decreases rapidly.
Remarkably, a $10$~m$^3$ detector employed to search for the luminous signal 
(and with a conservative assumption of the background as described above) has 
comparable sensitivity to what XENON100 has achieved by searching for 
the signal of inelastic nuclear recoil.

Larger volumes and/or lower backgrounds would significantly improve
the sensitivity of CYGNUS to the luminous signal, see \figref{fig:sigmaCYGNUS}. CYGNUS is designed as a directional dark matter detector and will suppress backgrounds by using pointing of the dark matter recoil signal.  The signal of luminous dark matter is not pointing but is instead modulating in a unique way.  As discussed for Borexino this feature also allows good suppression of backgrounds.
Since CYGNUS detector development
is still underway, there are opportunities in design and material
choices that are likely to reduce the background estimates.  
As an indication of the maximal possible reach we also show the sensitivity of a 
hypothetical CYGNUS-like $10$~m$^3$ gaseous detector with zero
background.  The improvement in the sensitivity --
over $100$ times better -- is highly significant since it would allow
the CYGNUS detector to achieve \emph{much} stronger sensitivity to
inelastic dark matter with mass splittings up to about 
$150$~keV\@.

\section{Discussion}

We demonstrated that we can use the Earth to inelastically upscatter dark matter 
into an excited state that later decays into a photon, providing an outstanding
opportunity for dark matter detection.  As shown in Figures~\ref{fig:sigma} and \ref{fig:sigmaCYGNUS},
large underground experiments including both large neutrino experiments 
such as Borexino and JUNO as well as large directional detection experiments 
such as CYGNUS can achieve sensitivities to the photon decay signal that are 
significantly stronger than inelastically scattering off xenon in a 
conventional direct detection experiment.  The improved sensitivities
benefit greatly from the large sidereal-daily modulation that provides
an excellent handle to separate signal from background.  Both trace 
abundance of heavy elements like lead as well as the much more 
numerous elements such as iron and silicon provide the main scattering nuclei.
This allowed us to obtain good estimates of the sensitivity of this technique, although we are still sensitive to
uncertainties in the elemental abundances.  More elements could be
included in calculating the sensitivities, and a more nuanced model of their distribution in the Earth, but these would not qualitatively
change our conclusions.  For instance, we estimate that including the 
heaviest 20 elements in the Earth improves our sensitivity by
less than a factor of $2$.
We emphasize that we have calculated Borexino's expected 
sensitivity, and not a limit.  Indeed, we encourage Borexino to 
re-analyze their data to look for a sidereal-daily modulating signal,
as they could undoubtedly improve upon the crude statistical measures
that we employed to obtain our estimates. 

We also found that large gaseous scintillation detectors such as CYGNUS 
are sensitive to the photon decay signal with smaller inelastic
splittings, $\delta \ltap 160$~keV, due to their lower detection
threshold and the absence of a $^{14}$C radioactive background.
In fact, CYGNUS could be more sensitive to the photon signal
at these lower inelastic splittings than the limits from 
inelastic nuclear recoil in existing xenon experiments.

In this paper our main focus was on dark matter that inelastically 
upscattered through one process (that could be due to $Z$-exchange)
and decayed via a radiative decay.  The notion that the upscatter
process $\chi_1 + N \rightarrow \chi_2 + N$ is separate from the 
decay $\chi_2 \rightarrow \chi_1 + \gamma$ occurs generically in a 
wide variety of models, including narrowly-split higgsinos
of split Dirac supersymmetry.  In the narrowly-split higgsino model, the ordinary 
elastic scattering process does occur at one-loop, but is highly 
suppressed by the twist-2 suppression of the operator as well as an 
accidental cancellation among the diagrams.  

It is also possible that the inelastic scattering process and the
radiative decay proceed through the same operator.  Magnetic
inelastic dark matter \cite{Chang:2010en,Barello:2014uda} 
is a known example of this.
Derived bounds \cite{Bramante:2016rdh} on the magnetic inelastic 
interaction from XENON100 and PICO data require the characteristic 
decay length of the excited state be at least 
$\ell \gtap 10, 100, 1000$~m for $\delta \ltap 150, 100, 50$~keV\@.  
The dark matter nucleus scattering in magnetic inelastic models involves several
operators (in the notation of \cite{Barello:2014uda}). 
It was shown in \cite{Bramante:2016rdh} that for lower $\delta$, 
the spin-independent interaction dominates, while for larger 
$\delta$, the spin-dependent interaction dominates.  This means 
that to determine the sensitivity of CYGNUS to magnetic inelastic
dark matter requires a careful treatment of the Earth's abundances of 
not only massive elements but also ones with a high spin.
We will present this analysis in future work \cite{ustoappear}.

It is interesting to consider if other detectors could be sensitive
to inelastic dark matter, possibly in other ranges of inelastic splittings.
SNO+ \cite{Andringa:2015tza} could be sensitive to a similar region of parameter space
that we have presented for Borexino and JUNO\@.  We did not calculate the
sensitivity of SNO+ due to its latitude  
(most of the Cygnus constellation
is always above the horizon at Sudbury) and its energy threshold being between
$200$-$400$~keV\@.  It is an accident of circumstance 
(where deep mining for nickel in Canada happens to be located)
that suggests SNO+ would have significantly poorer sensitivity
compared to JUNO, where in its location much higher rates 
are expected.  DUNE \cite{Acciarri:2016ooe} also provides an interesting possibility 
for a large underground experiment that could be sensitive to
a photon signal.  
Unfortunately, its threshold is higher than 
an MeV, and so it does not appear to be useful for galactic dark matter whose
maximum speed is limited by the escape velocity of the 
galaxy.  
It is clear that the luminous signal of inelastic dark matter provides an additional, and orthogonal, search technique with which to hunt for dark matter.  This ``photon phrontier'' may provide reach beyond that of the cross section or inelastic frontiers.

\section*{Acknowledgments}

We thank M.~Pospelov for being M.~Pospelov (pointing out relevant
physics in obscure sources).  We thank R.~Lang and N.~Spooner for 
helpful discussion, and important information on XENON100 and CYGNUS, respectively.
GDK thanks the Universities Research Association for 
travel support, and Fermilab for hospitality, where part of 
this work was completed.
The work of JE was supported by the Zuckerman STEM Leadership Program.  
The work of PJF and RH was supported by the DoE under contract number DE-SC0007859 and Fermilab, operated by Fermi Research Alliance, LLC under contract number DE-AC02-07CH11359 with the United States Department of Energy.
The work of GDK was supported in part by the
U.S. Department of Energy under Grant Number DE-SC0011640.

\bibliographystyle{utphys}
\bibliography{borexino}

\end{document}